\documentclass[12pt]{iopart}
\usepackage{iopams}
\usepackage{graphicx}
\expandafter\let\csname equation*\endcsname\relax
\expandafter\let\csname endequation*\endcsname\relax
\usepackage{amsmath}
\usepackage{amssymb}
\usepackage{cite}

\begin{document}

\title[First return times of random walks on random regular graphs]
{Analytical results for the distribution of  
first return times of random walks on random regular graphs
}

\author{Ido Tishby, Ofer Biham and Eytan Katzav}
\address{Racah Institute of Physics, 
The Hebrew University, Jerusalem 9190401, Israel.}
\eads{\mailto{ido.tishby@mail.huji.ac.il}, \mailto{biham@phys.huji.ac.il}, 
\mailto{eytan.katzav@mail.huji.ac.il}}

\begin{abstract}

We present analytical results for the distribution of
first return (FR) times of
random walks (RWs) on random regular graphs (RRGs)
consisting of $N$ nodes of degree $c \ge 3$.
Starting from a random initial node $i$ at time $t=0$, 
at each time step $t \ge 1$ an RW hops 
into a random neighbor of its previous node.
We calculate the distribution 
$P ( T_{\rm FR} = t )$
of first return times to the initial node $i$. 
We distinguish between first return trajectories in which the RW
retrocedes its own steps backwards all the way back to the initial node $i$
and those in which the RW returns to $i$ via a path that does not retrocede its own steps.
In the retroceding scenario, each edge that belongs to the RW trajectory is crossed the
same number of times in the forward and backward directions.
In the non-retroceding scenario
the subgraph that consists of the nodes visited by the RW and
the edges it has crossed between these nodes
includes at least one cycle.
In the limit of $N \rightarrow \infty$ the RRG converges towards the Bethe lattice.
The Bethe lattice exhibits a tree structure,
in which all the first return trajectories belong to the retroceding scenario.
Moreover, in the limit of $N \rightarrow \infty$ the trajectories of RWs on RRGs
are transient in the sense that they return to the initial node with probability $<1$.
In this sense they resemble the trajectories of RWs on regular lattices of dimensions
$d \ge 3$.
The analytical results are
found to be in excellent agreement with the results 
obtained from computer simulations.

\end{abstract}


\noindent{\it Keywords}: 
Random network, 
random regular graph,
random walk, 
first passage time,
first return time,
recurrence,
transience.

\maketitle

\section{Introduction}

Random walk (RW) models
\cite{Lawler2010}
provide useful tools for the analysis of
dynamical processes on random networks 
\cite{Havlin2010,Newman2010}
such as the
spreading of rumours, opinions and infections
\cite{Pastor-Satorras2001,Barrat2012,Masuda2017}.
Starting at time $t=0$ from a random initial node $i=x_0$, 
at each time step $t \ge 1$ the RW hops
randomly to one of the neighbors of its previous node.
The resulting trajectory takes the form
$x_0 \rightarrow x_1 \rightarrow \dots \rightarrow x_t \rightarrow \dots$,
where $x_{t}$ is the node visited at time $t$.
In some of the time steps the RW visits nodes that
have not been visited before, while
in other time steps it visits nodes that have
already been visited at an earlier time.
The mean number $\langle S \rangle_t$ of distinct nodes visited by an RW 
on a random network up to time $t$ was recently studied 
\cite{Debacco2015}.
It was found that 
in the infinite network limit
at sufficiently long times 
$\langle S \rangle_t \simeq r t$, 
where the coefficient
$r<1$ depends on the network topology.
These scaling properties resemble those obtained
for RWs on high dimensional lattices and Cayley trees
\cite{Masuda2004}.
They imply that RWs on random networks 
revisit previously visited nodes
less frequently than RWs on low dimensional lattices
\cite{Montroll1965}.
Therefore, RW models provide a highly effective framework for
search and exploration processes on random networks.

For an RW starting from an initial node $i$, the
first return (FR) time $T_{\rm FR}$ 
is the first time at which the RW 
returns to $i$
\cite{Redner2001}.
The first return time varies between different instances of the RW
trajectory and its properties can be captured by a suitable distribution.
The distribution of first return times may depend on the 
specific realization of the random network and on the
choice of the initial node $i$.
The distribution of first return times for a given ensemble of
random networks is denoted by $P(T_{\rm FR}=t)$.
This distribution is calculated by averaging over many
network instances drawn from the ensemble.
For each network instance one needs to sample many RW trajectories 
starting from random initial nodes.

A more general problem involves the calculation of
the first passage (FP) time $T_{\rm FP}$,
which is the first time at which an RW starting from an initial node $i$
visits a specified target node $j$
\cite{Redner2001,Sood2005,Peng2021}.
The first return problem is a special case of the first passage problem, 
in which the initial node is also chosen as the target node.
The distribution $P(T_{\rm FR}=t)$ of first return times was studied on the
Bethe lattice, which exhibits a tree structure of an infinite size
\cite{Hughes1982,Cassi1989,Giacometti1995}.
However, no closed-form analytical results are available for 
$P(T_{\rm FR}=t)$
on random networks
of a finite size $N$.

\begin{figure}
\centerline{
\includegraphics[width=4.4cm]{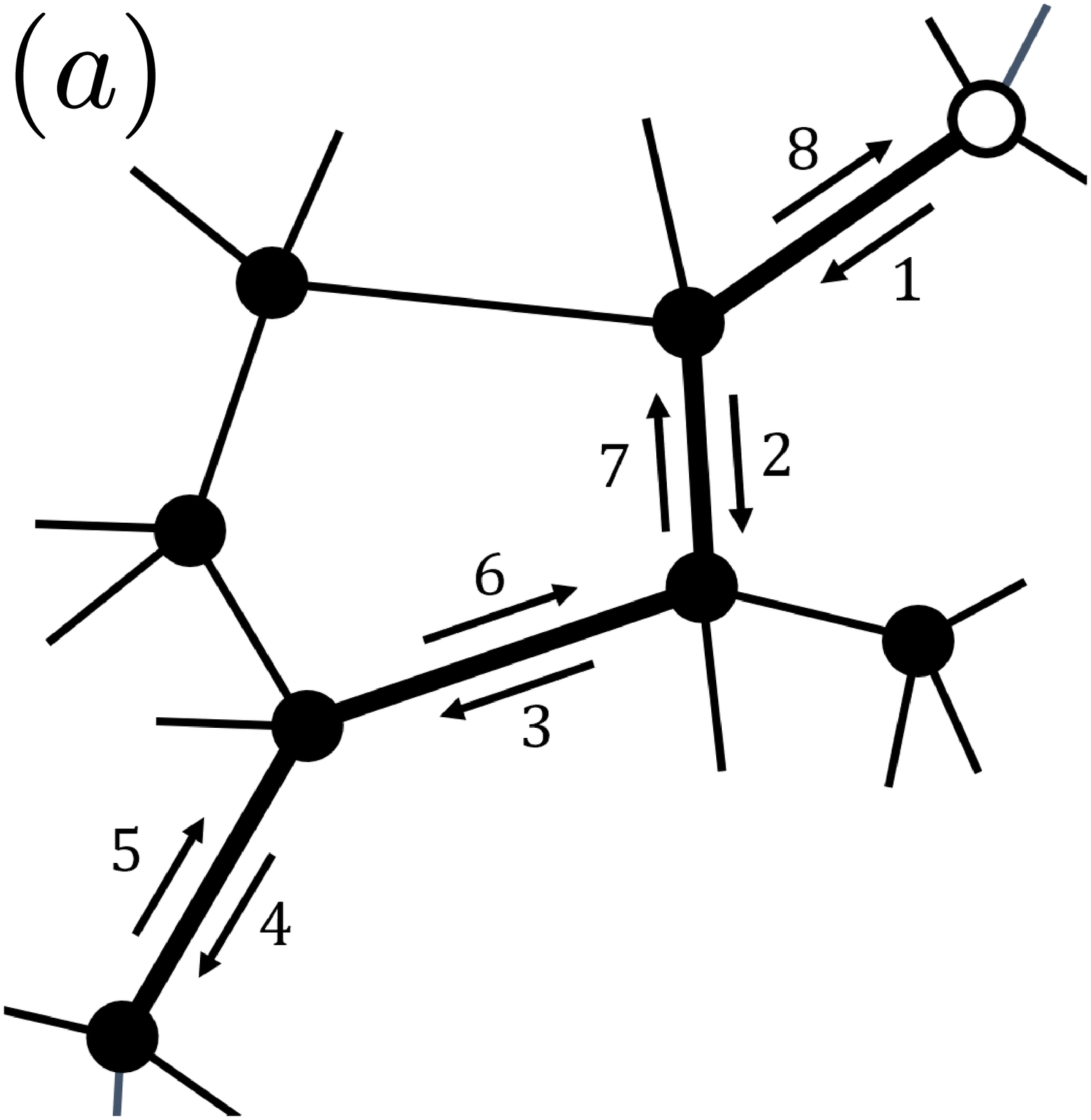}
}
\vspace{0.4in}
\centerline{
\includegraphics[width=4.4cm]{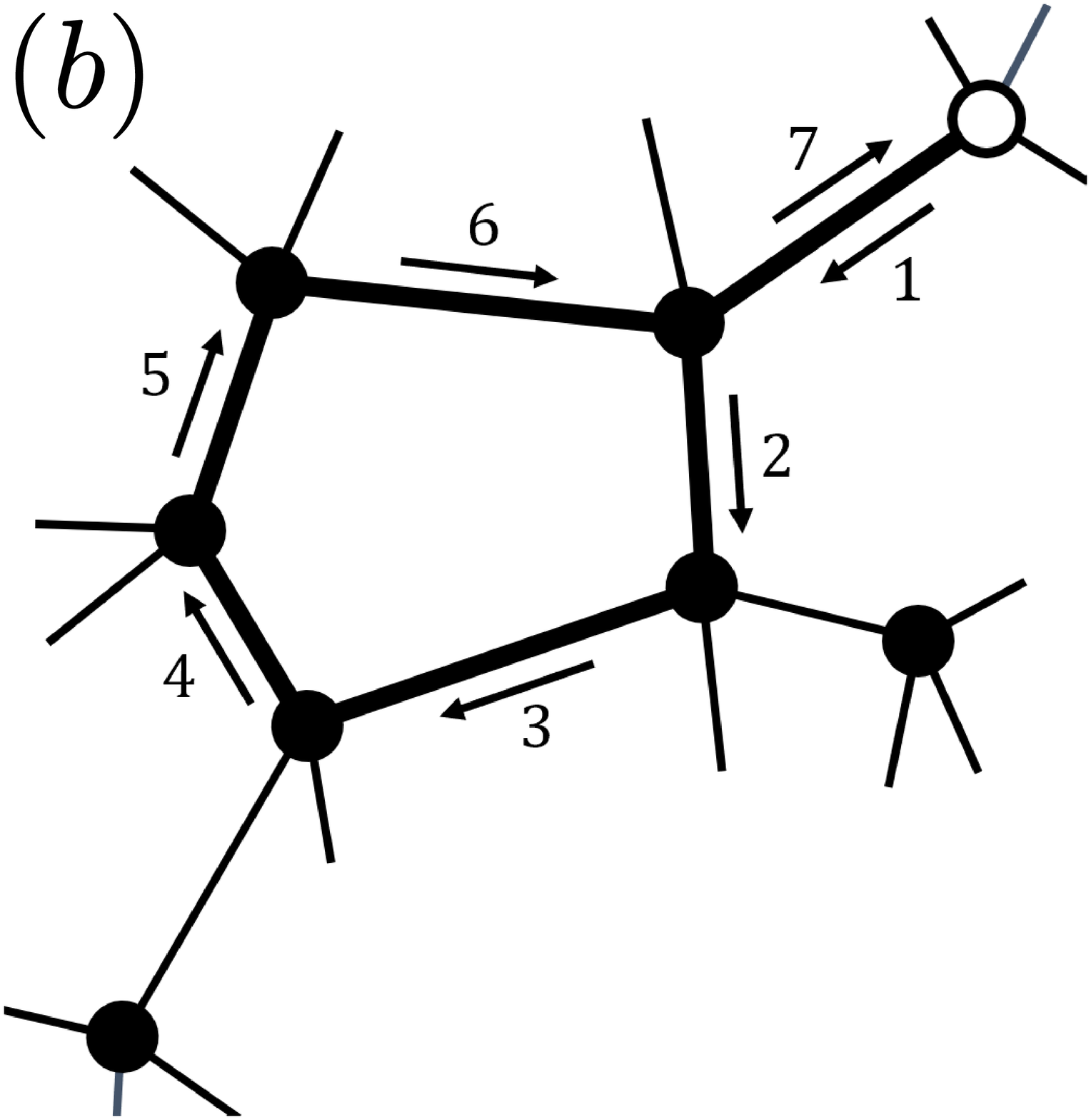}
}
\caption{
Schematic illustrations of a retroceding first-return trajectory (a)
and a non-retroceding first return trajectory (b) of an RW on an RRG.
Note that in this illustration the RRG is of degree $c=4$.
}
\label{fig:1}
\end{figure}

In this paper we present analytical results for 
the distribution of  
first return times 
of RWs on random regular graphs (RRGs) consisting of $N$ nodes of degree $c \ge 3$.
We consider
separately the scenario in which the RW returns
to the initial node $i$ by retroceding (RETRO) its own steps and
the scenario in which it does not retrocede its steps 
($\lnot {\rm RETRO}$) 
on the way back to $i$.
In the retroceding scenario an RW starting from the initial node $i$
forms a random trajectory in the network and eventually returns
to $i$ by stepping backwards via the same edges that it crossed
in the forward direction [Fig. \ref{fig:1}(a)]. This implies that in the retroceding scenario each edge
that belongs to the RW trajectory is crossed the same number of
times in the forward and backward directions (which means that the total number
of crossings of each edge must be an even number).
As a result, in the retroceding scenario the first return time must be even, namely
$T_{\rm FR}=2,4,6,\dots$.
In the infinite network limit, in which the RRG exhibits a
tree structure, all the first return trajectories belong to the retroceding scenario.
In this case the subgraph consisting of the nodes visited by the
RW and of the edges it has crossed between these nodes exhibits a tree structure.
Note that in an RRG of a finite size  
an RW may return at time $t'$ to a node that has been visited at an earlier time
$t < t'-2$, forming a cycle of length $t'-t$. 
There is a very rare possibility in which
the RW may then retrocede its steps, crossing all the edges along the cycle in the
backward direction to the node visited at time $t$ and then to the initial 
node $i$. 
In this case, the path is counted as a retroceding (RETRO) first return trajectory,
in spite of the fact that it includes a cycle.

In the non-retroceding scenario an RW starting from $i$ forms a random trajectory in
the network and eventually returns to $i$ without retroceding its own steps [Fig. \ref{fig:1}(b)].
This means that in the non-retroceding scenario
the trajectory includes at least three edges that are crossed
an odd number of times.
In the non-retroceding scenario the RW trajectory must include at least one cycle.
Therefore, this scenario takes place only in RRGs of a finite size and is diminished
in the infinite network limit.
In the non-retroceding scenario the first return time may take any even or odd
value that satisfies $T_{\rm FR} \ge 3$.
While the results for the retroceding trajectories are exact, 
in the analysis of the non-retroceding trajectories we use
approximations that are accurate in the large $N$ and long time limits.
Therefore, the analytical results for the non-retroceding scenario are
approximate results.

From the discussion above, it is clear that the retroceding and non-retroceding
scenarios are mutually exclusive. More specifically, a first return RW trajectory in which every
single edge is crossed the same number of times in the forward and backward 
directions belongs to the retroceding scenario, otherwise it 
belongs to the non-retroceding scenario. 
Using combinatorial and probabilistic methods we calculate the
conditional distributions of first return times,
$P(T_{\rm FR}=t | {\rm RETRO})$
and
$P(T_{\rm FR}=t | \lnot {\rm RETRO})$,
in the retroceding and non-retroceding scenarios, respectively.
We also calculate the mean and variance of each one of the two conditional distributions.
We combine the results of the two scenarios with suitable weights and obtain
the overall distribution of first return times
$P(T_{\rm FR} = t)$
of RWs on RRGs of a finite size.
In the limit of $N \rightarrow \infty$ the RRG converges towards the Bethe lattice.
The Bethe lattice exhibits a tree structure,
in which all the first return trajectories belong to the retroceding scenario.
It is also found that in the infinite network limit the trajectories of RWs on RRGs
are transient in the sense that they return to the initial node with probability $<1$.
In this sense they resemble the trajectories of RWs on regular lattices of dimensions
$d \ge 3$.
The analytical results are found to be in excellent agreement with the results 
obtained from computer simulations.

The paper is organized as follows.
In Sec. 2 we briefly describe the random regular graph.
In Sec. 3 we present the random walk model.
In Sec. 4 we calculate the distribution of first return times of RWs on RRGs
in the retroceding scenario.
In Sec. 5 we calculate the distribution of first return times 
in the non-retroceding scenario.
To this end, we derive a closed-form expression for $\langle S \rangle_t$ in RRGs
of a finite size, which is accurate at intermediate and long times.
In Sec. 6 we combine the results
obtained in sections 4 and 5 to obtain the overall distribution of first return times.
In Sec. 7 we use the results obtained for the distribution of first return times to 
derive a more refined expression for
$\langle S \rangle_t$, 
which is accurate also at short times.
In Sec. 8 we discuss the results and compare
the return probability of an RW on an RRG to the return probability of an RW on
a regular lattice with the same coordination number.
The results are summarized in Sec. 9.
In Appendix A we present an asymptotic expansion of P\'olya's constant,
which provides the return probability of an RW on a
$d$-dimensional hypercubic lattice.

\section{The random regular graph}

A random network (or graph) consists of a set of $N$ nodes that
are connected by edges in a way that is determined by some
random process.
For example, in a configuration model network the degree of each node is 
drawn independently from a given degree distribution $P(k)$ and
the connections are random and uncorrelated
\cite{Molloy1995,Molloy1998,Newman2001}.
The RRG is a special case of a configuration 
model network, in which the degree distribution is a degenerate
distribution of the form 
$P(k)=\delta_{k,c}$, namely
all the nodes are of the same degree $c$.
Here we focus on RRGs of a finite size $N$ and degree $c \ge 3$,
which for a sufficiently large $N$ consist of a single connected component
\cite{Bollobas2001}.

In the infinite network limit RRGs with a finite degree $c \ge 3$ exhibit a tree
structure with no cycles. 
Thus, in this limit it coincides with a Bethe lattice whose coordination number is equal to $c$.
In contrast, RRGs of a finite size exhibit a local tree-like structure,
while at larger scales there is a broad spectrum of cycle lengths.
In that sense RRGs differ from Cayley trees, which maintain their
tree structure by reducing the most peripheral nodes to leaf nodes of degree $1$.

A convenient way to construct an RRG
of size $N$ and degree $c$
($Nc$ must be an even number)
is to prepare the $N$ nodes such that each node is 
connected to $c$ half edges or stubs
\cite{Newman2010}.
At each step of the construction, one connects a random pair of stubs that 
belong to two different nodes $i$ and $j$ 
that are not already connected,
forming an edge between them.
This procedure is repeated until all the stubs are exhausted.
The process may get stuck before completion in case that
all the remaining stubs belong to the
same node or to pairs of nodes that are already connected.
In such case one needs to perform some random reconnections
in order to complete the construction.

\section{The random walk model}

\begin{figure}
\centerline{
\includegraphics[width=5.0cm]{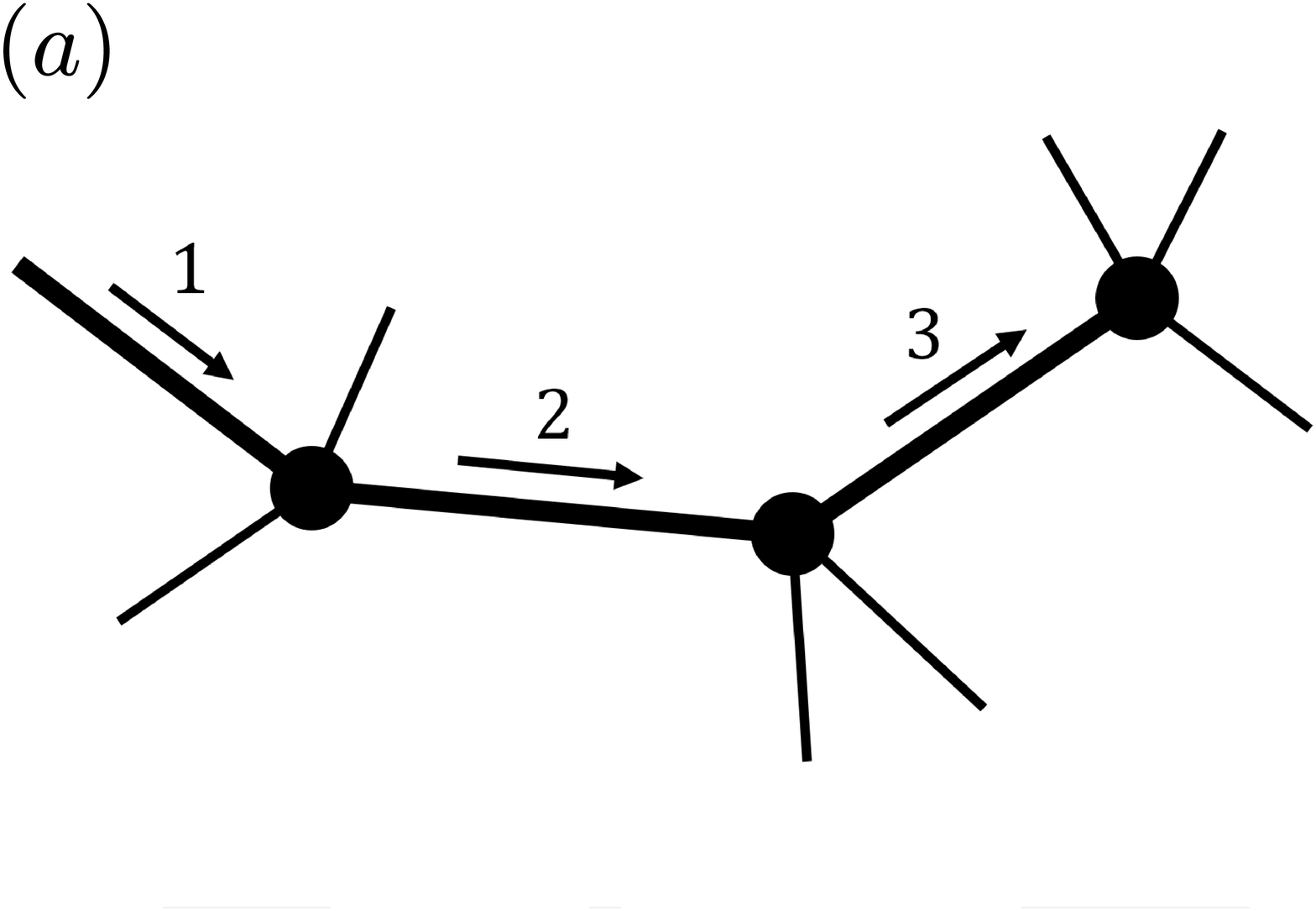}
}
\centerline{
\includegraphics[width=5.0cm]{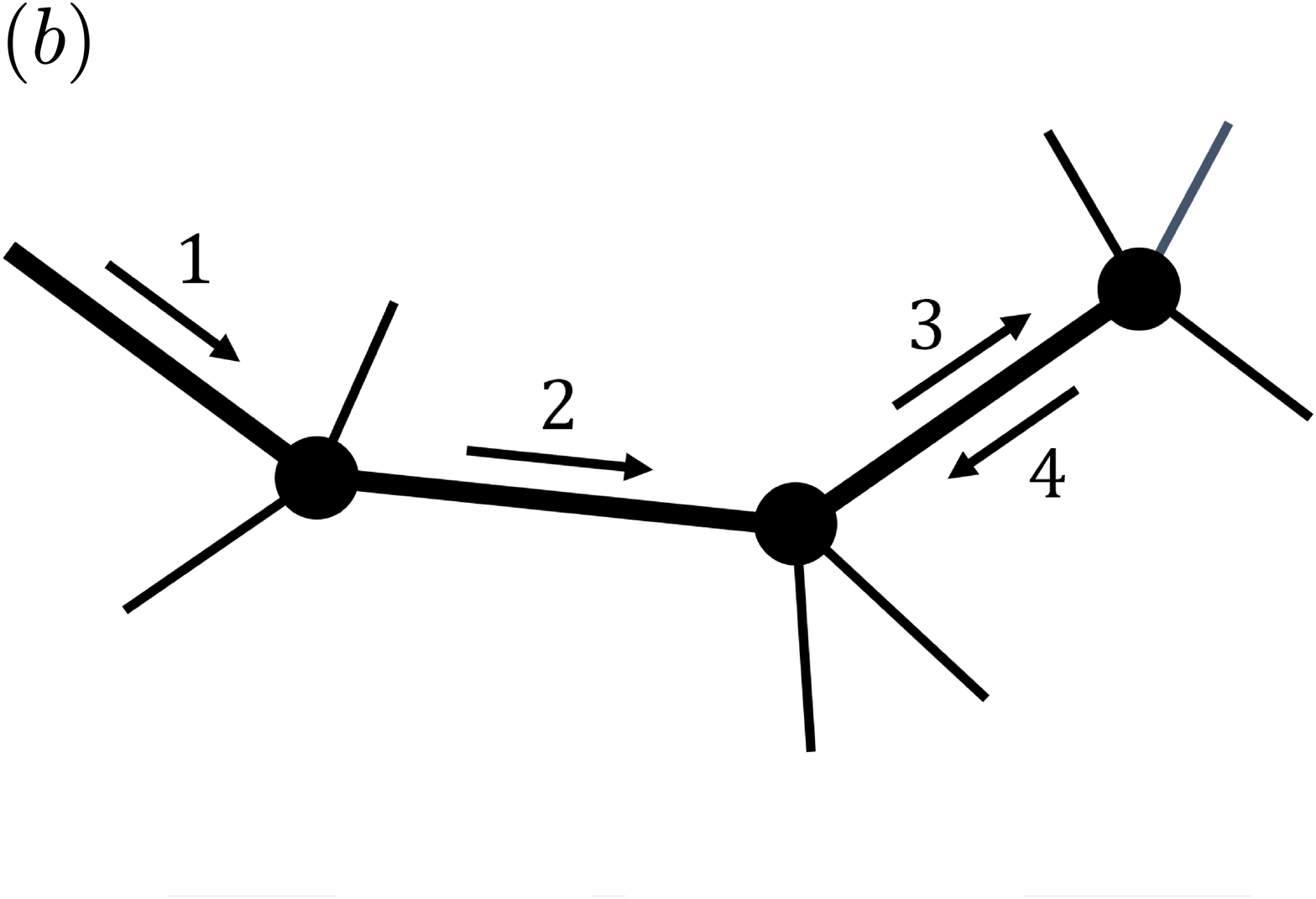}
}
\centerline{
\includegraphics[width=5.0cm]{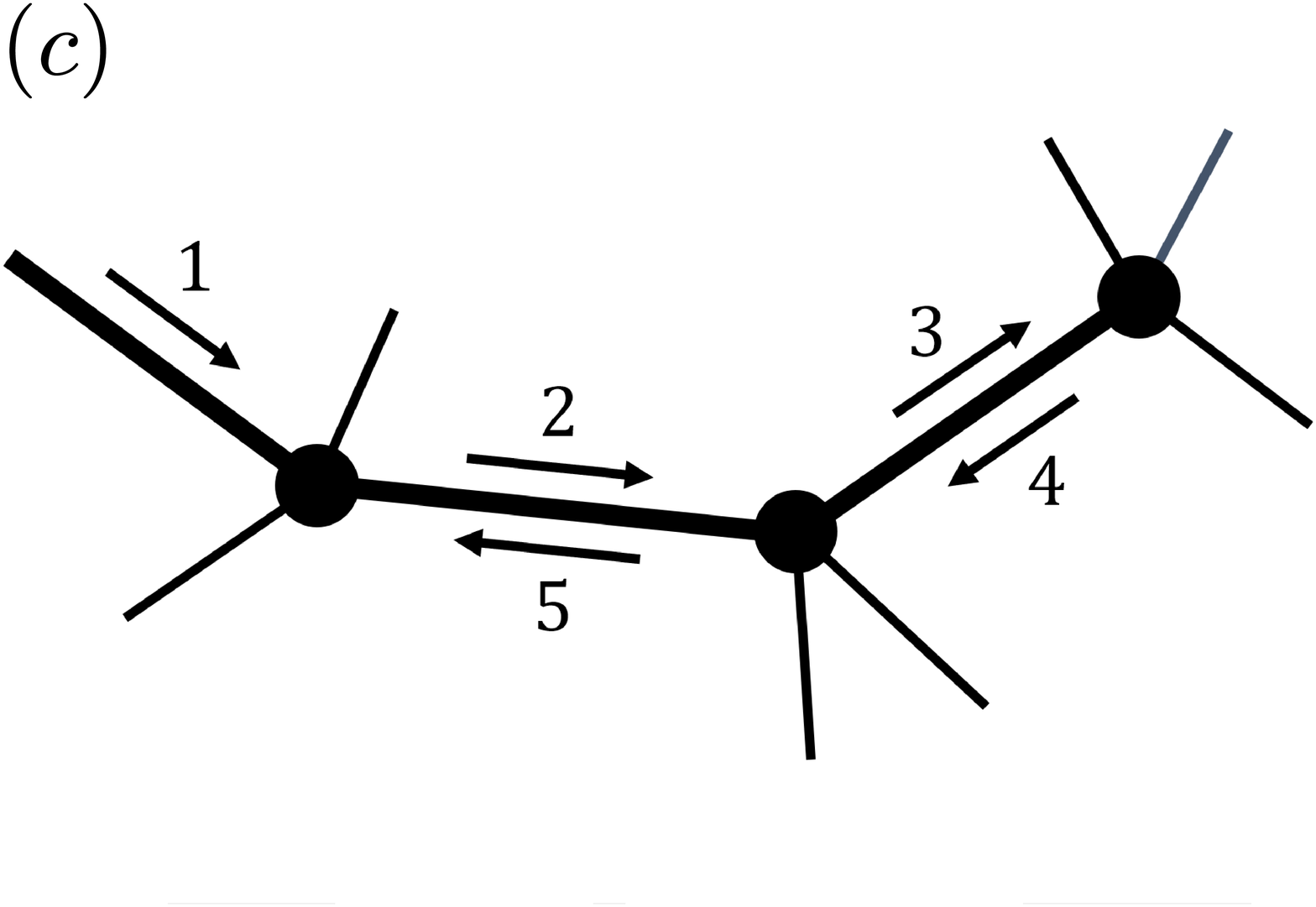}
}
\centerline{
\includegraphics[width=5.0cm]{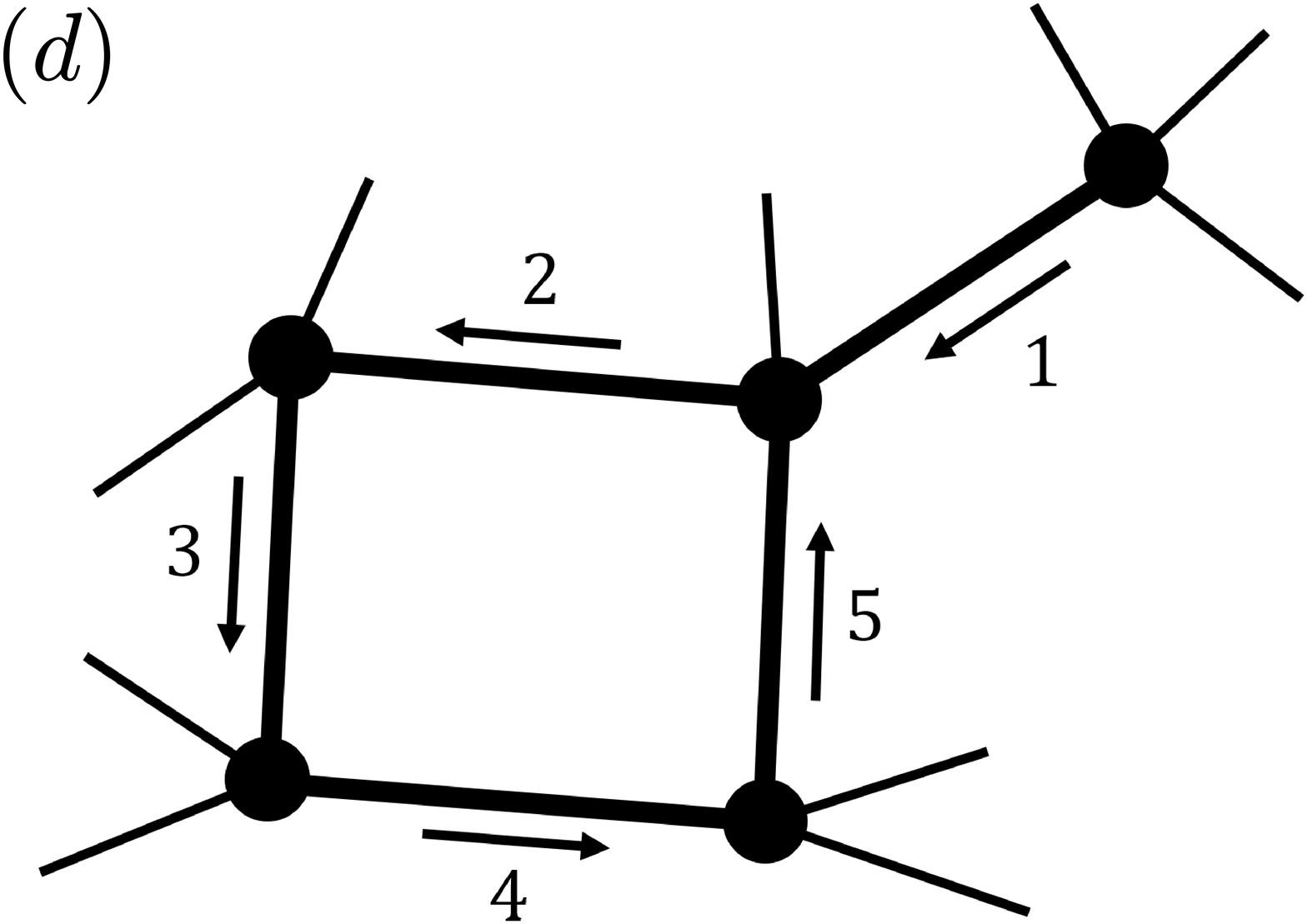}
}
\caption{
Schematic illustrations of possible events taking place along the path 
of an RW on an RRG:
(a) a path segment in which at each time step the RW enters a node
that has not been visited before; 
(b) a path segment that includes a backtracking step
into the previous node (step no. 4);
(c) a path segment that includes a backtracking step (step no. 4), which is 
followed by a retroceding step (step no. 5);
(d) a path that includes a retracing step (step no. 5) in which the RW hops into a node that was
visited a few time steps earlier. Retracing steps are not possible in the infinite
network limit and take place only in finite networks, which include cycles. 
Note that in this illustration the RRG
is of degree $c=4$.
}
\label{fig:2}
\end{figure}

Consider an RW on an RRG, 
starting from a random initial node $i$ at time $t=0$. 
At each time step $t \ge 1$ the RW hops 
randomly into one of the neighbors of its previous node.
Since RRGs with $c \ge 3$ consist of a single connected
component, an RW starting from any initial node
can reach any other node in the network.
At each time step $t \ge 2$ the RW may either step into a yet-unvisited node or
into a node that has already been visited before.
In Fig. \ref{fig:2} we present a
schematic illustration of some of the events that may take place along the path 
of an RW on an RRG.
In Fig. \ref{fig:2}(a) we show a path segment in which at each time step the RW enters a node
that has not been visited before.
In Fig. \ref{fig:2}(b) we show a path segment that includes a backtracking step,
in which the RW moves back into the previous node (step no. 4). 
In Fig. \ref{fig:2}(c) we show a path segment that includes a backtracking step
(step no. 4)
which is followed by a retroceding step (step no. 5).
In Fig. \ref{fig:2}(d) we show a path segment that includes a retracing step (step no. 5),
in which the RW enters a node that was visited four time steps earlier. 
Retracing steps are not possible in the infinite network limit in which the RRG
exhibits a tree structure.

The mean number of distinct nodes 
that are visited by an RW up to time $t$ 
is denoted by
$\langle S \rangle_t$.
The difference

\begin{equation}
\Delta_t = \langle S \rangle_{t} - \langle S \rangle_{t-1}
\label{eq:Delta_t}
\end{equation}

\noindent
is the probability that at time $t$ the RW will visit a node
that has not been visited before. 
For example, $\Delta_1=1$ and $\Delta_2=(c-1)/c$.
Using a generating function formulation based on the cavity method,
it was shown that 
the probability that an RW on an RRG of size $N \rightarrow \infty$
at time $t \gg 1$
will step into a yet-unvisited node 
is given by
\cite{Debacco2015}

\begin{equation}
\Delta_t = \frac{c-2}{c-1}.
\label{eq:Delta_tc}
\end{equation}

\noindent
A similar result was obtained for RWs on
the Bethe-Lattice  
\cite{Cassi1989,Martin2010}.
The behavior of $\Delta_t$ in RRGs of a finite size has not been studied.
However, 
Eq. (\ref{eq:Delta_tc}) 
provides accurate results for $\Delta_t$ 
on finite RRGs in a range of intermediate times $1 \ll t \ll N$, 
namely at times which are not too short but much shorter than
the network size.

An RW that returns to the initial node with probability $1$ is called a recurrent RW,
while an RW that returns to the initial node with probability $<1$ is called a transient RW
\cite{Spitzer2001}.
Recurrent RWs return to the initial node infinitely many times while transient RWs return
to the initial node only a finite number of times.
In a seminal paper by G. P\'olya, published 100 years ago,
it was shown that an RW on a $d$-dimensional hypercubic lattice is recurrent in 
dimensions $d=1,2$ and transient in dimensions $d \ge 3$
\cite{Polya1921}.
Thus, in general, RWs on infinite systems may be either recurrent or transient,
depending on the structure and dimension of the underlying network or lattice.
In contrast, RWs on finite systems are always recurrent.

\section{The distribution of first return times via retroceding trajectories}

Consider an RW on an RRG, starting at $t=0$ from a random initial node $i$.
In the infinite network limit the RW may either return to $i$ via a retroceding trajectory
or drift away without ever returning to $i$.
In a finite network the RW returns to $i$ with probability $1$ either via a retroceding
trajectory or via a non-retroceding trajectory.
We first consider the retroceding trajectories.
The probability that an RW will 
first return to $i$ at time $t$,
via a retroceding trajectory,
is given by

\begin{equation}
P(T_{\rm FR} = t,{\rm RETRO}) = 
\frac{ B(t) }{c^t},
\label{eq:PTFRR}
\end{equation}

\noindent
where $B(t)$, $t \ge 2$, is the number of retroceding RW trajectories 
that return to the initial node $i$ for the first time at time $t$.
In the retroceding first return trajectories each edge is crossed the same number of times
in the forward and in the backward directions.
Therefore, retroceding trajectories exist only for even values of $t$.
The first step is always an outward step (from $i$ to one of its neighbors),
while the last step is always an inward step (from one of the neighbors back to $i$).
The other $t-2$ steps can be ordered in many different ways, as long as at each intermediate
time the number of outward steps exceeds the number of inward steps by at least $1$.
The number of ways to order the inward and outward steps under this condition is given
by the Catalan number $C_{\frac{t}{2}-1}$,
where

\begin{equation}
C_{ k } = \frac{ (2k)! }{k! \left( k+1 \right)!  } 
= \frac{1}{2k+1} \binom{2k+1}{k+1} 
= \frac{1}{k+1} \binom{2k}{k}.
\label{eq:Catalan}
\end{equation}

\noindent
The Catalan number appears in many combinatorial problems.
For example, it counts
the number of discrete mountain ranges
of length $2k$
\cite{Audibert2010,Koshy2009}.
For an RRG of degree $c$ the number of possibility for the first (outward) step is $c$
while the number of possibilities for each one of the remaining $t/2-1$ outward steps
is $c-1$. 
The inward steps follow the path carved by the outer steps so they do not contribute
additional factors of $c$.
Multiplying the Catalan number by the factors of $c$ and $c-1$,
we obtain

\begin{equation}
B(t) = 
\left\{
\begin{array}{ll}
C_{ \frac{t}{2} - 1 } \ c (c-1)^{ \frac{t}{2} - 1 }   & \ \ \ \ \ \ \  t   \ \   {\rm even} \\
0 & \ \ \ \ \ \ \  t   \ \  {\rm odd}.
\end{array}
\right.  
\label{eq:Bt0}
\end{equation}

The overall probability that an RW will first return to its initial node $i$ 
via a retroceding trajectory is given by

\begin{equation}
P( {\rm RETRO} ) = 
\sum_{t=2}^{\infty}
P(T_{\rm FR}=t, {\rm RETRO}).
\label{eq:Pret}
\end{equation}

\noindent
Inserting $P(T_{\rm FR}=t, {\rm RETRO})$ from Eq. (\ref{eq:PTFRR})
into Eq. (\ref{eq:Pret}) and carrying out the summation, we obtain

\begin{equation}
P( {\rm RETRO} ) = 
\frac{1}{c-1}.
\label{eq:Pret2}
\end{equation}

\noindent
This result is in agreement with Eq. (30) in Ref. 
\cite{Hughes1982},
which provides the probability of return in the Bethe lattice.
Thus, in the infinite network limit, 
in which the RW is transient,
the probability $P({\rm R})$ that an RW will return
to its initial node $i$ satisfies $P({\rm R})=P({\rm RETRO})$.
The complementary probability that an RW on a finite RRG will first return to  
$i$ via a non-retroceding path 
(or that an RW on an infinite RRG will never return to $i$),
is given by

\begin{equation}
P( \lnot {\rm RETRO} ) = 
\frac{c-2}{c-1}.
\label{eq:Pret3}
\end{equation}

\begin{figure}
\centerline{
\includegraphics[width=6.8cm]{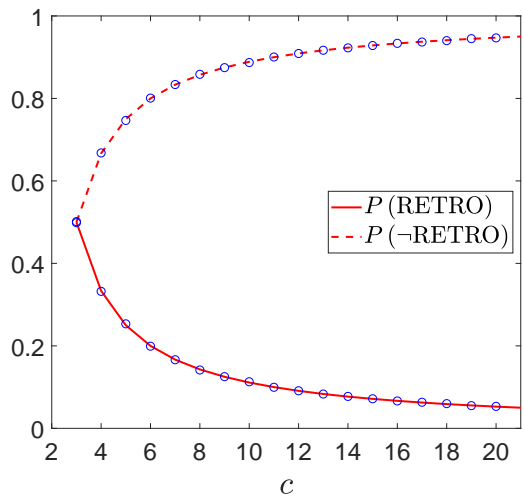}
}
\caption{
Analytical results for the
probability $P({\rm RETRO})$
that the first return process will take place 
via a retroceding trajectory
and the probability $P(\lnot {\rm RETRO})$
that it will take place via a non-retroceding trajectory,
as a function of the degree 
$c$ for random regular graphs 
of size $N=1000$.
The analytical results,
obtained from Eqs.
(\ref{eq:Pret2}) and (\ref{eq:Pret3}), respectively, 
are in excellent agreement with 
the results obtained from computer simulations (circles).
}
\label{fig:3}
\end{figure}

In Fig. \ref{fig:3} we present
analytical results for the
probability $P({\rm RETRO})$
that the first return process will take place 
via a retroceding trajectory
and the probability $P(\lnot {\rm RETRO})$
that it will take place via a non-retroceding trajectory,
as a function of the degree 
$c$ for random regular graphs 
of size $N=1000$.
The analytical results,
obtained from Eqs.
(\ref{eq:Pret2}) and (\ref{eq:Pret3}), respectively, 
are in excellent agreement with 
the results obtained from computer simulations (circles).

For the simulations we generated a large number (typically 100) 
of random instances of the RRG 
consisting of $N$ nodes of degree $c$, using the procedure 
presented in Sec. 2. For each network instance, we generated
a large number (typically 100) of RW trajectories, where each trajectory starts 
from a random initial node $i$ at time $t=0$. 
The simulation results are obtained by averaging over all these trajectories.
In the simulations, at each time step $t$ the RW selects randomly  
one of the $c$ neighbors of the node $x_{t-1}$,
where the probability of each neighbor to be selected 
is $1/c$. It then hops to the selected node, denoted by $x_t$.
Each RW trajectory is terminated upon its first return to $i$.
The first return time $t$ is thus equal to the length of the trajectory.
The trajectory $x_0,x_1,\dots,x_t$ is recorded for further analysis.
It is then determined whether the first return occurred via the
retroceding or the non-retroceding scenario.
More specifically, if all the edges along the RW trajectory
are crossed an equal number of times in the forward and the
backward directions
(where the forward direction is the direction of the first crossing),
we conclude that the first return occurred via
the retroceding scenario. Otherwise, we conclude that it occurred 
via the non-retroceding trajectory.

The distribution of first return times under the condition of retroceding paths
is given by

\begin{equation}
P(T_{\rm FR}=t | {\rm RETRO}) 
=
\frac{ P(T_{\rm FR}=t, {\rm RETRO}) }{ P({\rm RETRO}) }.
\label{eq:PTfrtr1}
\end{equation}

\noindent
Inserting $P(T_{\rm FR}=t, {\rm RETRO})$ from Eq. (\ref{eq:PTFRR}) 
and $P({\rm RETRO})$ from Eq. (\ref{eq:Pret2}) into Eq. (\ref{eq:PTfrtr1}),
we obtain

\begin{equation}
P(T_{\rm FR} = t | {\rm RETRO}) = 
\left\{
\begin{array}{ll}
C_{ \frac{t}{2} - 1 } 
\left( 1 - \frac{1}{c} \right)^{\frac{t}{2}} \left( \frac{1}{c} \right)^{\frac{t}{2}-1}  
&   \ \ \ \ \ \   t   \ \    {\rm even} \\
0 & \ \ \ \ \ \  t   \ \ \  {\rm odd},
\end{array}
\right.  
\label{eq:PTFRt2}
\end{equation}

\noindent
where $C_{ \frac{t}{2} - 1 }$ is given by Eq. (\ref{eq:Catalan}).
Eq. (\ref{eq:PTFRt2}) is in agreement with Eq. (24) in Ref. 
\cite{Masuda2004}, which provides the distribution of first return
times on Cayley trees.

The generating function of the distribution of first return times 
in the retroceding scenario
is denoted by

\begin{equation}
V_{\rm RETRO}(x) = \sum_{k=1}^{\infty}
x^{2k} P(T_{\rm FR}=2k | {\rm RETRO}).
\label{eq:V}
\end{equation}

\noindent
Inserting $P(T_{\rm FR}=2k | {\rm RETRO})$
from Eq. (\ref{eq:PTFRt2}) 
into Eq. (\ref{eq:V})
and carrying out the summation, we obtain

\begin{equation}
V_{\rm RETRO}(x) = \frac{ c - \sqrt{ c^2 - 4 (c-1)x^2 } }{2}.
\label{eq:V2}
\end{equation}

\noindent
The probability
$P(T_{\rm FR} = t | {\rm RETRO})$
can be obtained from the generating function 
by differentiation via

\begin{equation}
P(T_{\rm FR} = t | {\rm RETRO})=
\frac{1}{t!} \frac{d^t}{dx^t} V_{\rm RETRO}(x) \bigg\vert_{x=0}.
\end{equation}

\noindent
Inserting $x=1$ in Eq. (\ref{eq:V2})
we obtain $V(1)=1$, which confirms the normalization of
the distribution
$P(T_{\rm FR} = t | {\rm RETRO})$.

The tail distribution of first return times is given by

\begin{equation}
P(T_{\rm FR}>t| {\rm RETRO}) =
\sum_{t'=t+1}^{\infty} P(T_{\rm FR} = t'| {\rm RETRO}).
\label{eq:PTsum}
\end{equation}

\noindent
Inserting $P(T_{\rm FR} = t | {\rm RETRO})$ from Eq. (\ref{eq:PTFRt2})
into Eq. (\ref{eq:PTsum}) and carrying out the summation, we obtain

\begin{equation}
P(T_{\rm FR}>t| {\rm RETRO}) =
C_{ \lfloor \frac{t}{2} \rfloor } 
\frac{c (c-1)}{(c-2)^2}
\left( \frac{c-1}{c^2} \right)^{ \lfloor \frac{t}{2} \rfloor }
\, _2F_1 \left[ \left.
\begin{array}{c}
1,  \frac{3}{2} \\
\lfloor \frac{t}{2} \rfloor + 2
\end{array}
\right| - \frac{4(c-1)}{(c-2)^2} 
\right],
\label{eq:PTsum2}
\end{equation}

\noindent
where 
$\lfloor x \rfloor$ is the integer part of $x$.
The function
$_2F_1[ \ ]$
on the right hand side of Eq. (\ref{eq:PTsum2})
is the hypergeometric function,
which is given by

\begin{equation}
_2F_1 \left[ \left.
\begin{array}{c}
a, b \\
c
\end{array}
\right| z 
\right] =
\sum_{n=0}^{\infty} 
\frac{ (a)_n (b)_n }{ (c)_n } \frac{ z^n }{ n! },
\label{eq:2F1}
\end{equation}

\noindent
where $(q)_n$ is the (rising) Pochhammer symbol
\cite{Olver2010}.

In order to calculate the moments of 
$P(T_{\rm FR} = t | {\rm RETRO})$
we introduce the moment generating function 
$M_{\rm RETRO}(x)=V_{\rm RETRO}(e^x)$.
It is given by

\begin{equation}
M_{\rm RETRO}(x) = \frac{ c - \sqrt{ c^2 - 4 (c-1) e^{2x} } }{2}.
\label{eq:M}
\end{equation}

\noindent
Expanding $M_{\rm RETRO}(x)$ to second order in powers of $x$,
we obtain the mean first return time

\begin{equation}
\mathbb{E}[T_{\rm FR} | {\rm RETRO}] = \frac{ 2(c-1) }{c-2},
\label{eq:ETFRet}
\end{equation}

\noindent
and the second moment

\begin{equation}
\mathbb{E}[T_{\rm FR}^2 | {\rm RETRO}] = \frac{ 4(c-1)(c^2 - 2c +2 ) }{(c-2)^3}.
\end{equation}

\noindent
Combining the results presented above for the first and second moments,
we obtain the variance

\begin{equation}
{\rm Var}[T_{\rm FR} | {\rm RETRO}] = \frac{4 c(c-1)}{(c-2)^3}.
\label{eq:VarFRet}
\end{equation}

\noindent
From Eq. (\ref{eq:ETFRet}) we conclude that the mean first return time is
of order $1$, and from Eq. (\ref{eq:VarFRet}) we conclude that the
distribution of first return times is very narrow.
This means that the retroceding scenario 
takes place only at very short times.

\section{The distribution of first return times via non-retroceding trajectories}

Consider an RW starting from a random initial node $i$ at $t=0$.
The mean number of distinct nodes visited by the RW up to time $t$ 
is denoted by $\langle S \rangle_{t}$.
The probability $P(T_{\rm FR} \le t | \lnot {\rm RETRO})$ 
that the RW will return to the initial node up to time $t$ under the 
condition that it will occur via the non-retroceding scenario
is essentially the probability
that one of the 
$\langle S \rangle_{t-1}$ 
nodes visited during $t'=1,2,\dots,t$ is the initial node $i$.
This probability is given by

\begin{equation}
P(T_{\rm FR} \le t | \lnot {\rm RETRO}) =
\left\{
\begin{array}{ll}
0   & \ \ \ \ \ \   t=0,1,2  \\
\frac{\langle S \rangle_{t-1} - 2 }{N-2}   & \ \ \ \ \ \  t \ge 3.  \\
\end{array}
\right.
\label{eq:PTFRtR0}
\end{equation}

\noindent
The numerator on the right hand side of Eq.  (\ref{eq:PTFRtR0}),
for $t \ge 3$, represents the expected number of distinct nodes visited by an RW
in the time interval $1 \le t' \le t$, apart from the nodes
visited at $t'=1$ and $t'=2$. The point is that the nodes visited at times
$t'=1$ and $t'=2$ in the non-retroceding scenario
are clearly distinct from the initial node $i$.
In contrast, at any later time, $t' \ge 3$ the RW may return to $i$
via the non-retroceding scenario.
The denominator on the right hand side of Eq. (\ref{eq:PTFRtR0})
represents the number of nodes in the network, apart from
the nodes visited at $t'=1$ and $t'=2$, namely the same two
nodes that are offset in the numerator. While the RW may visit
these two nodes at times $t' \ge 3$, such visits would not make
any contribution to $\langle S \rangle_{t-1}$ because these would
be return visits.
In the RRG all the nodes are of the same degree while the connectivity is random.
Therefore, all the nodes in the network 
(apart from the nodes visited at $t'=1$ and $t'=2$)
have the same probability to be visited by the RW 
in the time interval $3 \le t' \le t$.
Therefore, the right hand side of Eq. (\ref{eq:PTFRtR0}) 
represents the probability that an RW will return to
the initial node $i$ up to time $t$. 
The proper normalization of this expression is
apparent from the fact that 
$(\langle S \rangle_{t-1} - 2)/(N-2) \rightarrow 1$
in the limit of $t \rightarrow \infty$.

The complementary probability, namely the probability that the RW will not return to $i$ within
the first $t$ time steps, 
under the condition that the first return process will take place via the
non-retroceding scenario,
is given by

\begin{equation}
P(T_{\rm FR}>t  | \lnot {\rm RETRO} ) =
\left\{
\begin{array}{ll}
1   & \ \ \ \ \ \   t=0,1,2  \\
1 - \frac{\langle S \rangle_{t-1} - 2 }{N-2}   & \ \ \ \ \ \  t \ge 3.  \\
\end{array}
\right.
\label{eq:P_FRT_tail}
\end{equation}

\noindent
In order to utilize Eq. (\ref{eq:P_FRT_tail}) we derive below a closed-form expression
for $\langle S \rangle_t$. 
To this end, we first consider the probability $\Delta_t$,
which at intermediate times $1 \ll t \ll N$ 
it is given by Eq. (\ref{eq:Delta_tc}).
At longer times, $\Delta_t$ is proportional to the
fraction of yet-unvisited nodes among all the nodes in the network,
apart from the nodes visited at times $t-1$ and $t-2$.
Therefore, a saturation term emerges and

\begin{equation}
\Delta_t = 
\left\{
\begin{array}{ll}
1   & \ \ \ \ \ \   t=1  \\
\frac{c-2}{c-1} \left( 1 - \frac{\langle S \rangle_{t-1}-2}{N-2} \right)    & \ \ \ \ \ \  t \ge 2.  \\
\end{array}
\right.
\label{eq:Delta_tS}
\end{equation}

\noindent
Eq. (\ref{eq:Delta_tS}) indicates that
the probability of an RW at time $t$ to enter a yet-unvisited node
depends on $t$ via the expected number of distinct nodes 
$\langle S \rangle_{t-1}$
that have been visited up to time $t-1$.
Thus, the probability of an RW that has already visited $s$ distinct
nodes to enter a yet-unvisited node in the next time step is given by

\begin{equation}
\Delta(s) = 
\left\{
\begin{array}{ll}
1   & \ \ \ \ \ \   s=1  \\
\frac{c-2}{c-1} \left( 1 - \frac{s-2}{N-2} \right)     & \ \ \ \ \ \  s \ge 2.  \\
\end{array}
\right.
\label{eq:Deltas}
\end{equation}

\begin{figure}
\centerline{
\includegraphics[width=7cm]{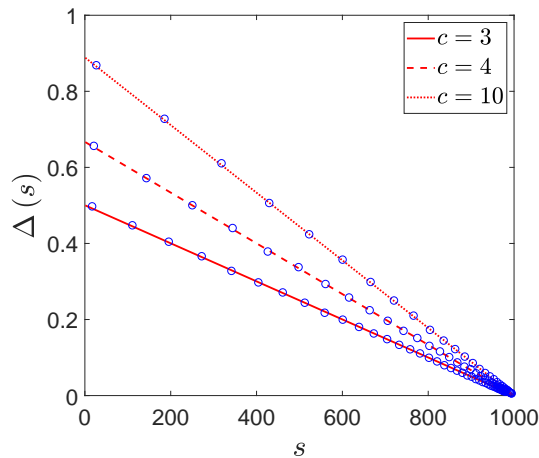}
}
\caption{
Analytical results for the
probability $\Delta(s)$ of an RW 
that has already visited $s$ distinct nodes
to hop into a new node 
that has not been visited before,
on a random regular graph of size $N=1000$ and degrees
$c=3$ (solid line), $c=4$ (dashed line)
and $c=10$ (dotted line).
The analytical results,
obtained from Eq. (\ref{eq:Deltas}),
are in excellent agreement with the results
obtained from computer simulations (circles).
} 
\label{fig:4}
\end{figure}

In Fig. \ref{fig:4} we present
analytical results for the
probability $\Delta(s)$ of an RW 
that has already visited $s$ distinct nodes
to hop in the next time step into a node 
that has not been visited before,
on RRGs of size $N=1000$ and degrees
$c=3$ (solid line), $c=4$ (dashed line)
and $c=10$ (dotted line).
The analytical results,
obtained from Eq. (\ref{eq:Deltas}),
are in excellent agreement with the results
obtained from computer simulations (circles).

Inserting $\Delta_t$ from Eq. (\ref{eq:Delta_tS}) into
Eq. (\ref{eq:Delta_t}), 
we obtain a recursion equation 
for $\langle S \rangle_t$
of the form

\begin{equation}
\langle S \rangle_{t}=
\frac{c-2}{c-1} \left( 1 + \frac{2}{N-2} \right)
+ \left[1- \left( \frac{c-2}{c-1} \right) \frac{1}{N-2} \right] \langle S \rangle_{t-1},
\label{eq:Steq}
\end{equation}

\noindent
that applies to $t \ge 2$.
Solving Eq. (\ref{eq:Steq}) with the initial condition 
$\langle S \rangle_1=2$,
we obtain

\begin{equation}
\langle S \rangle_{t}
= 
2 +
(N-2) \left\{ 1 - \left[1 - \left( \frac{ c-2 }{ c-1 } \right) \frac{1}{N-2} \right]^{t-1} \right\}. 
\label{eq:Stsol}
\end{equation}

\noindent
Approximating the
square brackets  
on the right hand side 
of Eq. (\ref{eq:Stsol})
by an exponential function,
we obtain

\begin{equation}
\langle S \rangle_{t}
\simeq 
2 + (N-2) \left\{ 1 - \exp \left[ -  \left( \frac{c-2}{c-1} \right)   \frac{t-1}{N-2}   \right] \right\}. 
\label{eq:Stlate2}
\end{equation}

\noindent
Eq. (\ref{eq:Stlate2}) depends on time only via $(t-1)/(N-2)$, 
which highlights the fact that $\langle S \rangle_t$ is a slowly varying quantity.
It can be characterized by the
time $\tau$ at which the RW is expected to complete visiting half of the nodes
in the network, or formally $\langle S \rangle_{\tau} = N/2$.
The time $\tau$, given by

\begin{equation}
\tau = \left( \frac{c-1}{c-2} \right)  N  \ln(2) ,
\end{equation}

\noindent
is analogous to the half-life time of  
radioactive materials, which is the time it takes until half of the 
nuclei undergo radioactive decay.

\begin{figure}
\centerline{
\includegraphics[width=7cm]{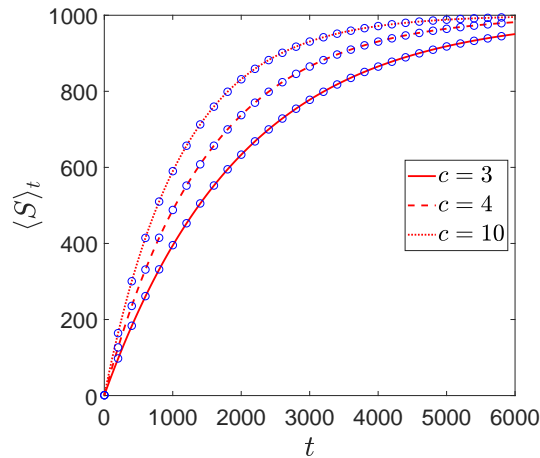}
}
\caption{
Analytical results for the
mean number of distinct nodes 
$\langle S \rangle_t$, 
visited by an RW 
up to time $t$ (solid lines)
on random regular graphs of size
$N=1000$
and degrees $c=3$ (solid line), $c=4$ (dashed line)
and $c=10$ (dotted line).
The analytical results,
obtained from Eq. (\ref{eq:Stlate2}),
are in excellent agreement with the results
obtained from computer simulations (circles).
} 
\label{fig:5}
\end{figure}

In Fig. \ref{fig:5} 
we present analytical results for
the mean number $\langle S \rangle_t$ 
of nodes visited by an RW up to time $t$
on an RRG of size
$N=1000$
and degrees 
$c=3$ (solid line), $c=4$ (dashed line)
and $c=10$ (dotted line).
The analytical results,
obtained from Eq. (\ref{eq:Stlate2}),
are in excellent agreement with the results
obtained from computer simulations (circles).
The results demonstrate the linear behavior
of $\langle S \rangle_t$ vs. $t$ at early times.
As $c$ is increased the
slope $d \langle S \rangle_t/dt$ at early times becomes steeper.
This implies that as $c$ is increased
the number of distinct nodes visited by the
RW up to a given time $t$ increases.

\begin{figure}
\centerline{
\includegraphics[width=7cm]{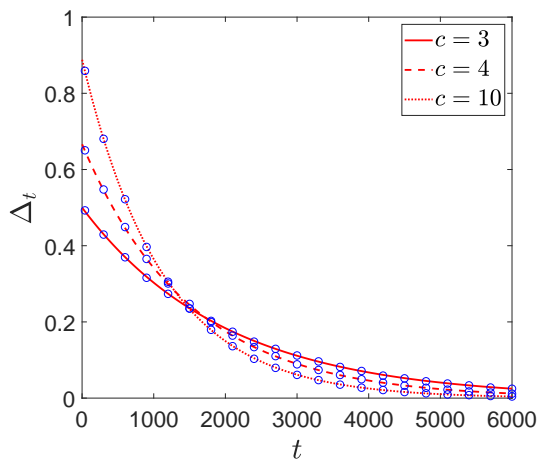}
}
\caption{
Analytical results for the
probability $\Delta_t$ of an RW at time $t$ to 
hop into a new node that has not been visited before,
on a random regular graph of size $N=1000$ and degrees
$c=3$ (solid line), $c=4$ (dashed line)
and $c=10$ (dotted line).
The analytical results,
obtained from Eq. (\ref{eq:Delta_tS}),
in which $\langle S \rangle_{t-1}$
is given by Eq. (\ref{eq:Stlate2}),
are in excellent agreement with the results
obtained from computer simulations (circles).
} 
\label{fig:6}
\end{figure}

In Fig. \ref{fig:6} 
we present analytical results 
for the probability $\Delta_t$ of an RW at time $t$ to 
step into a node that has not been visited before,
on an RRG of size $N=1000$ and degrees
$c=3$ (solid line), $c=4$ (dashed line)
and $c=10$ (dotted line).
The analytical results,
obtained from Eq. (\ref{eq:Delta_tS}),
in which $\langle S \rangle_{t-1}$
is given by Eq. (\ref{eq:Stlate2})
are in excellent agreement with the results
obtained from computer simulations (circles).

Plugging in $\langle S \rangle_t$ from 
Eq. (\ref{eq:Stlate2}) into Eq. (\ref{eq:P_FRT_tail})
and taking the long-time limit,
we obtain the tail distribution of first return times via the
non-retroceding scenario.
It is given by

\begin{equation}
P(T_{\rm FR} > t | \lnot {\rm RETRO}   ) = 
\left\{
\begin{array}{ll}
1   & \ \ \ \ \ \   t=0,1,2  \\
\exp \left[ -    \left( \frac{c-2}{c-1} \right)   \frac{t-2}{N-2}     \right]    & \ \ \ \ \ \  t \ge 3.  \\
\end{array}
\right.
\label{eq:P_FRT_tail2}
\end{equation}

\noindent
The probability mass function
of the distribution of first return times is given by

\begin{equation}
P(T_{\rm FR} = t | \lnot {\rm RETRO}   )=
P(T_{\rm FR} > t-1 | \lnot {\rm RETRO}   )
- P(T_{\rm FR} > t | \lnot {\rm RETRO}   ),
\end{equation}

\noindent
or

\begin{equation}
P(T_{\rm FR} = t | \lnot {\rm RETRO}   ) = 
\left\{
\begin{array}{ll}
0   & \ \   t=0,1,2  \\
\exp \left[ -    \left( \frac{c-2}{c-1} \right)   \frac{t-3}{N-2} \right]
-
\exp \left[ -    \left( \frac{c-2}{c-1} \right)   \frac{t-2}{N-2} \right]    & \ \  t \ge 3.  \\
\end{array}
\right.
\label{eq:P_FRT_tail2p}
\end{equation}

\noindent
The generating function of the distribution of first return times
in the non-retroceding scenario is given by

\begin{equation}
V_{\lnot {\rm RETRO}}(x) = 
\sum_{t=3}^{\infty} x^t P(T_{\rm FR}=t|\lnot {\rm RETRO}).
\label{eq:VnR}
\end{equation}

\noindent
Inserting $P(T_{\rm FR}=t|\lnot {\rm RETRO})$
from Eq. (\ref{eq:P_FRT_tail2p}) into Eq. (\ref{eq:VnR}), 
we obtain

\begin{equation}
V_{\lnot {\rm RETRO}}(x) =
x^2 \left\{ 1 + 
\frac{ x  - 1 }{1 - x \exp \left[ - \left( \frac{c-2}{c-1} \right) \frac{1}{N-2} \right] } \right\}.
\end{equation}

\noindent
In order to calculate the moments of $P(T_{\rm FR}=t|\lnot {\rm RETRO})$
we introduce the moment generating function
$M_{\lnot {\rm RETRO}}(x) = V_{\lnot {\rm RETRO}}(e^x)$.
It is given by

\begin{equation}
M_{\lnot {\rm RETRO}}(x) =
e^{2x} \left\{ 1 + 
\frac{ e^{x} - 1 }{1 - e^{x} \exp \left[ - \left( \frac{c-2}{c-1} \right) \frac{1}{N-2} \right] } \right\}.
\end{equation}

\noindent
We also introduce the cumulant generating function  
$K_{\lnot {\rm RETRO}}(x) = \ln M_{\lnot {\rm RETRO}}(x)$.
It is given by

\begin{equation}
K_{\lnot {\rm RETRO}}(x) =
2x + 
\ln \left\{ 1 + \frac{ e^{x} - 1 }{1 - e^{x} \exp \left[ - \left( \frac{c-2}{c-1} \right) \frac{1}{N-2} \right] } \right\}.
\end{equation}

\noindent
Expanding $K_{\lnot {\rm RETRO}}(x)$ in powers of $x$,
we obtain

\begin{eqnarray}
K_{\lnot {\rm RETRO}}(x) &=&
\left\{ 2 + \frac{ 1 }{1 -   \exp \left[ - \left( \frac{c-2}{c-1} \right) \frac{1}{N-2} \right] } \right\} x
\nonumber \\
&+&
\frac{ \exp \left[ - \left( \frac{c-2}{c-1} \right) \frac{1}{N-2} \right] }
{2 \left\{ 1 -   \exp \left[ - \left( \frac{c-2}{c-1} \right) \frac{1}{N-2} \right]  \right\}^2 } x^2
+ \mathcal{O} \left( x^3 \right).
\end{eqnarray}

\noindent
Thus, the mean first return time in the non-retroceding scenario is given by

\begin{equation}
\mathbb{E}[ T_{\rm FR} | \lnot {\rm RETRO} ]
=  2 +
\frac{1}{ 1 - \exp \left[ - \left( \frac{c-2}{c-1} \right) \frac{1}{N-2}   \right] },
\label{eq:t_FRT0}
\end{equation}

\noindent
and the variance of the distribution of first return times is given by

\begin{equation}
{\rm Var}[T_{\rm FR} | \lnot {\rm RETRO}] =  
\frac{   \exp \left[- \left( \frac{c-2}{c-1} \right) \frac{1}{N-2} \right]  }
{ \left\{ 1 - \exp \left[- \left( \frac{c-2}{c-1} \right) \frac{1}{N-2} \right] \right\}^2 }.
\label{eq:t_FPT_2}
\end{equation}

\noindent
Expanding these expressions for large $N$, we obtain

\begin{equation}
\mathbb{E} [ T_{\rm FR} | \lnot {\rm RETRO} ] \simeq  
\left( \frac{c-1}{c-2} \right) N + \frac{1}{2} - \frac{2}{c-2},
\label{eq:t_FRT}
\end{equation}

\noindent
and

\begin{equation}
{\rm Var}[T_{\rm FR} | \lnot {\rm RETRO}] \simeq  
\left( \frac{c-1}{c-2} \right)^2 (N-2)^2.
\label{eq:VarFR2}
\end{equation}

\noindent
Therefore, in the large $N$ limit both the mean and the
standard deviation of the distribution of first return times 
scale like $N$. This implies that
the distribution of first return times via
the non-retroceding scenario is a broad distribution.

\section{The overall distribution of first return times}
 
The overall distribution of first return times can be expressed in the form

\begin{eqnarray}
P(T_{\rm FR} = t) &=&
P(T_{\rm FR} = t| {\rm RETRO}) P( {\rm RETRO} ) 
\nonumber \\
&+&
P(T_{\rm FR} = t| \lnot {\rm RETRO}) P( \lnot {\rm RETRO} ).
\label{eq:PTFRgt}
\end{eqnarray}

\noindent
Since 
$\mathbb{E}[ T_{\rm FR} | {\rm RETRO} ]$
is of order $1$, while
$\mathbb{E}[ T_{\rm FR} | \lnot {\rm RETRO} ]$
is of order $N$, there is a clear separation of time scales
between the retroceding and the non-retroceding scenarios.
Inserting 
$P(T_{\rm FR} = t| {\rm RETRO})$
from Eq. (\ref{eq:PTFRt2})
and
$P(T_{\rm FR} = t| \lnot {\rm RETRO})$
from Eq. (\ref{eq:P_FRT_tail2p})
into Eq. (\ref{eq:PTFRgt}),
we obtain

\begin{equation}
P(T_{\rm FR} = t)  =  
\left\{
\begin{array}{ll}
\frac{B(t)}{c^t}   & \     t=0,1,2  \\
\frac{B(t)}{c^t} +
\left( \frac{c-2}{c-1} \right) \left\{ \exp \left[ -   \left( \frac{c-2}{c-1} \right)   \frac{t-3}{N-2}          \right]
- \exp \left[ -   \left( \frac{c-2}{c-1} \right)   \frac{t-2}{N-2}          \right] \right\}    & \    t \ge 3.  \\
\end{array}
\right.
\label{eq:PTFRgt2}
\end{equation}

\noindent
where $B(t)$ is given by Eq. (\ref{eq:Bt0}).
In the limit of $N \rightarrow \infty$, Eq. (\ref{eq:PTFRgt2}) 
is reduced to

\begin{equation}
P(T_{\rm FR} = t) =
\frac{B(t)}{c^t}.
\label{eq:PTFRgt7}
\end{equation}

\noindent
The generating function of $P(T_{\rm FR} = t)$
is given by

\begin{equation}
V(x) = \frac{ c - \sqrt{ c^2 - 4 (c-1)x^2 } }{2(c-1)}
+ x^2  \left\{ 1 +
\left( \frac{c-2}{c-1} \right)
\frac{x  - 1}
{1 - x \exp \left[ - \left( \frac{c-2}{c-1} \right) \frac{1}{N-2} \right] } \right\}.
\label{eq:Vtot}
\end{equation}

\noindent
The tail distribution of first return times can be expressed in the form

\begin{eqnarray}
P(T_{\rm FR} > t) &=&
P(T_{\rm FR} > t| {\rm RETRO}) P( {\rm RETRO} ) 
\nonumber \\
&+&
P(T_{\rm FR} > t| \lnot {\rm RETRO}) P( \lnot {\rm RETRO} ).
\label{eq:PTFRgt2p}
\end{eqnarray}

\noindent
Inserting
$P(T_{\rm FR} > t| {\rm RETRO})$
from Eq. (\ref{eq:PTsum2}) 
and
$P(T_{\rm FR} > t| \lnot {\rm RETRO})$
from Eq. (\ref{eq:P_FRT_tail2}) into Eq. (\ref{eq:PTFRgt2p}),
we obtain

\begin{eqnarray}
P(T_{\rm FR} > t) 
&=&
C_{ \lfloor \frac{t}{2} \rfloor }
\frac{c}{(c-2)^2}
\left(  \frac{c-1}{c^2} \right)^{ \lfloor \frac{t}{2} \rfloor }
\, _2F_1 \left[ \left.
\begin{array}{c}
1,  \frac{3}{2} \\
\lfloor \frac{t}{2} \rfloor + 2
\end{array}
\right|  -  \frac{4(c-1)}{(c-2)^2}
\right]
\nonumber \\
&+&
\left( \frac{c-2}{c-1} \right)
\exp \left[ -   \left( \frac{c-2}{c-1} \right) \frac{t-2}{N-2}        \right].
\label{eq:PTFRgt3}
\end{eqnarray}

\noindent
In the limit of $N \rightarrow \infty$ 
Eq. (\ref{eq:PTFRgt3}) is reduced to

\begin{eqnarray}
P(T_{\rm FR} > t) 
&=&
C_{ \lfloor \frac{t}{2} \rfloor }
\frac{c}{(c-2)^2}
\left(  \frac{c-1}{c^2} \right)^{ \lfloor \frac{t}{2} \rfloor }
\, _2F_1 \left[ \left.
\begin{array}{c}
1,  \frac{3}{2} \\
\lfloor \frac{t}{2} \rfloor + 2
\end{array}
\right|  -  \frac{4(c-1)}{(c-2)^2}
\right]
\nonumber \\
&+&
\left( \frac{c-2}{c-1} \right).
\label{eq:PTFRgt4}
\end{eqnarray}

The mean first return time $\langle T_{\rm FR} \rangle$
of RWs on RRGs can be obtained exactly using the Kac lemma,
which employs general properties of discrete stochastic processes
\cite{Kac1947}.
Applying the Kac lemma to an RW on an undirected random graph
consisting of a single connected component of size $N$, it implies that
the mean first return time to a given initial node $i$ is  
$\langle T_{\rm FR} \rangle_i = 1/P(i)$, 
where $P(i)$ is the fraction of time at which the RW resides at node $i$
in the limit of an infinitely long trajectory. 
In general, $P(i)$ is given by
$P(i) = k_i/[N \langle K \rangle$],
where $k_i$ is the degree of $i$ and $\langle K \rangle$
is the mean degree of the network
\cite{Masuda2017}.
In the case of an RRG consisting of $N$ nodes,
$P(i)=1/N$ for all the nodes.
As a result, the mean first return time for all nodes is
$\langle T_{\rm FR} \rangle = N$.

\begin{figure}
\centerline{
\includegraphics[width=14cm]{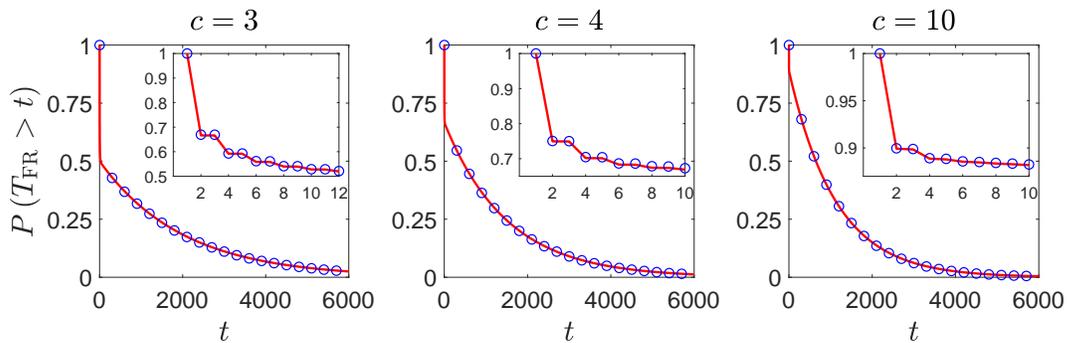}
}
\caption{
Analytical results for the 
tail distribution  
$P(T_{\rm FR} > t)$ (solid lines)
of first return times
of an RW on a random regular graph 
of size $N=1000$ 
and degrees
$c=3$, $c=4$ 
and $c=10$.
The insets show a magnified view of 
$P(T_{\rm FR} > t)$
at very short times, where the retroceding scenario is dominant.
The analytical results,
obtained from Eq.
(\ref{eq:PTFRgt3}),
are in excellent agreement with the results
obtained from computer simulations (circles).
}
\label{fig:7}
\end{figure}

Combining the results obtained above for the
retroceding and the non-retroceding scenarios, the
mean first return time
$\langle T_{\rm FR} \rangle$
can be expressed in the form

\begin{eqnarray}
\langle T_{\rm FR} \rangle &=&
\mathbb{E}[T_{\rm FR}  | {\rm RETRO}] P( {\rm RETRO} ) 
\nonumber \\
&+&
\mathbb{E}[T_{\rm FR} | \lnot {\rm RETRO}] P( \lnot {\rm RETRO} ).
\label{eq:<TFR>}
\end{eqnarray}

\noindent
Inserting
$\mathbb{E}[T_{\rm FR}  | {\rm RETRO}]$
from Eq. (\ref{eq:ETFRet})
and
$\mathbb{E}[T_{\rm FR} | \lnot {\rm RETRO}]$
from Eq. (\ref{eq:t_FRT0}) into Eq. (\ref{eq:<TFR>}),
we obtain

\begin{equation}
\langle T_{\rm FR} \rangle =
\frac{2}{c-2} + \left( \frac{c-2}{c-1} \right)
\left\{ 2 + \frac{1}{ 1 - \exp \left[ - \left( \frac{c-2}{c-1} \right) \frac{1}{N-2} \right] } \right\}.
\label{eq:LTfrR}
\end{equation}

\noindent
In the large network limit of $N \gg 1$, we obtain

\begin{equation}
\langle T_{\rm FR} \rangle = N + \mathcal{O} \left( 1 \right).
\end{equation}

\noindent
This result essentially agrees with the Kac lemma, apart from a discrepancy
of order $1$, which does not scale with $N$. 
The discrepancy is due to the approximations used in the derivation of the
distribution of first return times in the non-retroceding scenario, in which we
took the limits of large $N$ and long times.

In Fig. \ref{fig:7} 
we present analytical results for the tail distribution 
of first return times of RWs on an RRG  
of size $N=1000$ and 
$c=3$, $c=4$ 
and $c=10$.
The analytical results,
obtained from  Eq. (\ref{eq:PTFRgt3}),  
are in excellent agreement with the results
obtained from computer simulations (circles).

\begin{figure}
\centerline{
\includegraphics[width=7.2cm]{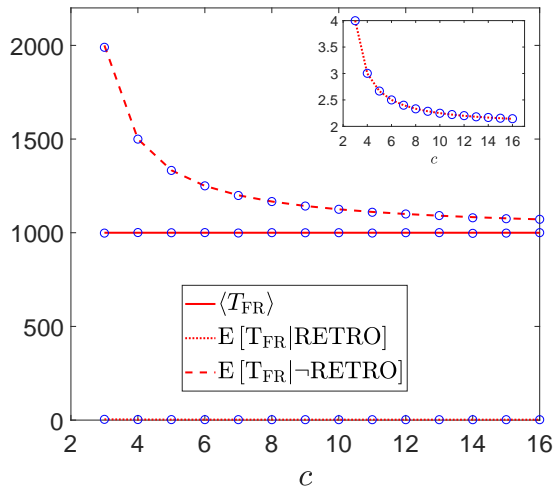}
}
\caption{
Analytical results for the
mean first return time 
$\langle T_{\rm FR} \rangle$ (solid line),
the mean first return time 
$\mathbb{E}[T_{\rm FR}|{\rm RETRO}]$
conditioned on
retroceding trajectories
(dotted line),
and the mean first return time 
$\mathbb{E}[T_{\rm FR}|\lnot {\rm RETRO}]$
conditioned on non-retroceding trajectories (dashed line),
as a function of the degree 
$c$ for random regular graphs 
of size $N=1000$.
The inset shows a magnified view of
$\mathbb{E}[T_{\rm FR}|{\rm RETRO}]$,
revealing the $c$-dependence of this curve.
The analytical results,
obtained from Eqs.
(\ref{eq:LTfrR}), 
(\ref{eq:ETFRet})
and (\ref{eq:t_FRT0}),  
are in excellent agreement with 
the results obtained from computer simulations (circles).
}
\label{fig:8}
\end{figure}

In Fig. \ref{fig:8} 
we present analytical results for the
mean first return time 
$\langle T_{\rm FR} \rangle$ (solid line),
the mean first return time conditioned on
retroceding trajectories
$\mathbb{E}[T_{\rm FR}|{\rm RETRO}]$ (dotted line)
and on non-retroceding trajectories 
$\mathbb{E}[T_{\rm FR}|\lnot {\rm RETRO}]$
(dashed line),
as a function of the degree 
$c$ for random regular graphs 
of size $N=1000$.
The analytical results,
obtained from Eqs.
(\ref{eq:LTfrR}), 
(\ref{eq:ETFRet})
and (\ref{eq:t_FRT0}), respectively,
are in excellent agreement with 
the results obtained from computer simulations (circles).

\section{A more accurate analytical expression for $\langle S \rangle_t$}

Consider an RW path segment of the form 
$x_0 \rightarrow x_1 \rightarrow x_2 \dots \rightarrow x_t$ 
on an RRG consisting of $N$ nodes of degree $c$.
For any initial node $x_0$, the probability of a given path
of length $t$ to occur is $1/c^t$.
Thus, in an infinitely long RW trajectory, the path segment
$x_0 \rightarrow x_1 \rightarrow x_2 \dots \rightarrow x_t$
(referred to as the forward path)
would occur with the same frequency as the time reversal (or backward) path
$x_t \rightarrow x_{t-1} \rightarrow x_{t-2} \rightarrow x_1 \rightarrow x_0$
in which the RW visits the same sequence of nodes but in the opposite direction.

Below we use an ensemble of such path segments to explore the
relation between the distribution of first return times $P(T_{\rm FR}>t)$ 
and the probability $\Delta_t$.
The probability $P(T_{\rm FR}>t)$ that the FR time of an RW starting from
a random node $x_0$ will be larger than $t$ is equal to
the fraction of forward path segments that satisfy
$x_{t'} \ne x_0$ for $t' = 1,2,\dots,t$,
drawn from an ensemble of RW path segments.
Similarly, the probability $\Delta_t$ that an RW will step into a previously
unvisited node at time $t$ is equal to the fraction of forward path segments
that satisfy $x_{t} \ne x_{t'}$ for $t'=0,1,2,\dots,x_{t-1}$.
Since the statistical weights of the backward path segments are equal
to the weights of the corresponding forward path segments, 
the probability $P(T_{\rm FR}>t)$ is also equal to the fraction of
backward path segments that satisfy
$x_t \ne x_{t'}$ for $t'=t-1,t-2,\dots,0$.
Similarly, $\Delta_t$ is equal to the fraction of backward path segments
that satisfy $x_0 \ne x_{t'}$ for $t'=t,t-1,\dots,1$.
Combining these results, we conclude that

\begin{equation}
\Delta_t = P(T_{\rm FR}>t).
\label{eq:DeltaPFR}
\end{equation}

In the limit of $N \rightarrow \infty$ the probability
$P(T_{\rm FR}>t)$ 
is given by
Eq. (\ref{eq:PTFRgt4}).
Using Eqs. (\ref{eq:PTFRgt4}) and (\ref{eq:DeltaPFR}) 
we evaluate $\Delta_t$ for the first few steps of the RW:

\begin{equation}
\Delta_t =
\left\{
\begin{array}{ll}
1  & \ \ \ \ \ \   t=0,1   \\
\frac{c-1}{c}  & \ \ \ \ \ \  t=2,3 \\
\frac{ (c-1)^2 (c+1)}{c^3}  & \ \ \ \ \ \  t=4,5 \\
\frac{ (c-1)^3 (c^2+2c+2) }{c^5}  & \ \ \ \ \ \  t=6,7.
\end{array}
\right.  
\label{eq:shorttimes}
\end{equation}

\noindent
In general, closed form expressions for $\Delta_t$ can be
obtained using the recursion equations

\begin{equation}
\Delta_t = 
2 \left( \frac{c-1}{c^2} \right)
\left\{ \left[ \frac{4 \lfloor t/2 \rfloor -3}{\lfloor t/2 \rfloor} + \frac{ (c-2)^2 }{2(c-1)} \right] \Delta_{t-2}
- \frac{2 \lfloor t/2 \rfloor -3}{\lfloor t/2 \rfloor} \Delta_{t-4} \right\},
\end{equation}

\noindent
where the initial conditions 
are given by Eq. (\ref{eq:shorttimes}).
This equation is obtained from Eq. (\ref{eq:PTFRgt4}).
More specifically, the initial conditions are 
$\Delta_0$ and $\Delta_2$ for even times
and $\Delta_1$ and $\Delta_3$ for odd times.
For $t \ge 2$ the probability $\Delta_t$ can also be expressed in the form

\begin{equation}
\Delta_t = c \left( \frac{c-1}{c^2} \right) Q_{\lfloor t/2 \rfloor - 1}(c),
\end{equation}
 
\noindent
where $Q_n(c)$ is a polynomial of order $n$ in $c$.
These polynomials can be generated by the recursion equation

\begin{equation}
Q_{k-1}(c) = 
\left[ \frac{4k-6}{k} + \frac{c^2}{(c-1)} \right] Q_{k-2}(c)
- \left( \frac{4k-6}{k} \right) \frac{c^2}{c-1} Q_{k-3}(c),
\end{equation}

\noindent
where $Q_0(c)=1$ and $Q_1(c)=c+1$.

The generating function associated with the probability $\Delta_t$ is given by

\begin{equation}
W(x) = \sum_{k=0}^{\infty}
x^{2k} \Delta_{2k}.
\end{equation}

\noindent
Note that the sum includes only the even terms due to the fact that
$\Delta_{2k+1}=\Delta_{2k}$.
Carrying out the summation, we obtain

\begin{equation}
W(x) = \frac{c-2 + \sqrt{c^2 - 4(c-1)x^2}}{2(c-1)(1-x^2)}.
\end{equation}

\noindent
The probability $\Delta_{2k}$ can be obtained from the generating function
according to

\begin{equation}
\Delta_{2k} = \frac{1}{(2k)!} \frac{d^{2k}}{dx^{2k}} W(x) \bigg\vert_{x=0}.
\end{equation}

\noindent
In the long time limit of $t \gg 1$ the hypergeometric function 
in Eq. (\ref{eq:PTFRgt4})
converges towards $1$.
Therefore, in this limit

\begin{equation}
\Delta_t 
 \simeq 
\frac{ c  }{\sqrt{\pi} (c-1)(c-2)^2  \big\lfloor \frac{t}{2} \big\rfloor^{ \frac{3}{2} } }
\left[ \frac{4(c-1)}{c^2} \right]^{  \lfloor \frac{t}{2}  \rfloor }
+
  \frac{c-2}{c-1}.  
\label{eq:PTFRgt6}
\end{equation}

\noindent
The first term on the right hand side of Eq. (\ref{eq:PTFRgt6}) 
decays exponentially with a characteristic time scale of

\begin{equation}
\tau = \frac{2}{ \ln \left[ \frac{c^2}{4(c-1)} \right] }.
\end{equation}

\noindent
As a result  

\begin{equation}
\Delta_t \rightarrow \frac{c-2}{c-1},
\label{eq:Dasympt}
\end{equation}

\noindent
in agreement with Eq. (\ref{eq:Delta_tc}).

The mean number of distinct nodes visited by an RW on an RRG up to time $t$
can be expressed in the form

\begin{equation}
\langle S \rangle_t =  \sum_{t'=0}^{t} \Delta_{t'}.
\label{eq:Sdelta}
\end{equation}

\noindent
Replacing $\Delta_t$ in
Eq. (\ref{eq:Sdelta}) 
by
$P(T_{\rm FR}>t)$
and performing summation by parts
\cite{Olver2010},
we obtain

\begin{equation}
\langle S \rangle_t = 
(t+1) P(T_{\rm FR}>t) 
+ \sum_{t'=1}^t t' P(T_{\rm FR}=t').
\label{eq:Sdelta2}
\end{equation}

\noindent
Inserting 
$P(T_{\rm FR}>t)$ from Eq. (\ref{eq:PTFRgt4})
$P(T_{\rm FR}=t')$ from Eq. (\ref{eq:PTFRgt7}) 
into Eq. (\ref{eq:Sdelta2}),
we obtain

\begin{eqnarray}
\langle S \rangle_t &=& 
2 +
C_{ \lfloor \frac{t}{2} \rfloor }
\frac{c (t+1)}{(c-2)^2}
\left(  \frac{c-1}{c^2} \right)^{ \lfloor \frac{t}{2} \rfloor }
\, _2F_1 \left[ \left.
\begin{array}{c}
1,  \frac{3}{2} \\
\lfloor \frac{t}{2} \rfloor + 2
\end{array}
\right|  -  \frac{4(c-1)}{(c-2)^2}
\right]
\nonumber \\
&+&
\frac{2}{c-1} 
\left[ - 1 +
\sum_{k=1}^{\lfloor t/2 \rfloor}
k C_{k-1}
\left( 1 - \frac{1}{c} \right)^k 
\left( \frac{1}{c} \right)^{k-1}
\right]
+ \frac{c-2}{c-1} (t-1).
\label{eq:Sdelta3}
\end{eqnarray}

\noindent
Carrying out the summation, we obtain

\begin{eqnarray}
\langle S \rangle_t &=& 
2 +
C_{  \lfloor \frac{t}{2}  \rfloor }
\frac{c (t+1)}{(c-2)^2}
\left(  \frac{c-1}{c^2} \right)^{  \lfloor \frac{t}{2}  \rfloor }
\, _2F_1 \left[ \left.
\begin{array}{c}
1,  \frac{3}{2} \\
\lfloor \frac{t}{2} \rfloor + 2
\end{array}
\right|  -  \frac{4(c-1)}{(c-2)^2}
\right]
\nonumber \\
&-&
2
C_{ \lfloor \frac{t}{2} \rfloor }
\frac{c \left(\lfloor \frac{t}{2} \rfloor + 1 \right)}{(c-2)^2}
\left(  \frac{c-1}{c^2} \right)^{ \lfloor \frac{t}{2} \rfloor }
\, _2F_1 \left[ \left.
\begin{array}{c}
1,  \frac{1}{2} \\
 \lfloor \frac{t}{2}  \rfloor + 1
\end{array}
\right|  -  \frac{4(c-1)}{(c-2)^2}
\right]
\nonumber \\
&+& 
\frac{2}{(c-1)(c-2)} +
\frac{c-2}{c-1}  (t-1).
\label{eq:Sdelta4}
\end{eqnarray}

\noindent
In the long time limit,  
the terms in Eq. (\ref{eq:Sdelta4}) that include the hypergeometric functions
decay to zero. As a result, Eq. (\ref{eq:Sdelta4}) is reduced to

\begin{equation}
\langle S \rangle_t \simeq
2 +
\frac{2}{(c-1)(c-2)} +
\frac{c-2}{c-1}  (t-1).
\label{eq:Sdelta5}
\end{equation}

\noindent
In the case of finite networks, 
in the long-time limit,
the last equation can be extended to the form

\begin{equation}
\langle S \rangle_{t}
\simeq 
2 +  
(N - 2) \left\{ 1 - \exp \left[ - \frac{2}{(c-1)(c-2)(N-2)}-  \left( \frac{c-2}{c-1} \right)   \frac{t-2}{N-2}     \right] \right\}. 
\label{eq:Stlate7}
\end{equation}

\noindent
This equation is more accurate than Eq. (\ref{eq:Stlate2}),
because it takes into account the contribution of the first few steps to $\langle S \rangle_t$
in an exact way. The essence of this is that at early times 
the discovery rate $\Delta_t$ of new, yet-unvisited nodes,
given by Eq. (\ref{eq:shorttimes}),
is slightly larger
than its asymptotic value, given by Eq. (\ref{eq:Dasympt}).

\section{Discussion}

First passage processes are an important landmark in the life-cycle of
RWs on networks. The characteristic time scale of these processes is
of order $t \sim N$.  
Another landmark is the first hitting process, which is the first time
in which the RW enters a previously visited node
\cite{Tishby2017,Tishby2021}.
The characteristic time scale of the first hitting process is 
$t \sim \min \{c,\sqrt{N} \}$,
namely $t \sim c$ in dilute networks and 
$t \sim \sqrt{N}$ in dense networks.
In both cases the first hitting time is much shorter than the first passage time
\cite{Tishby2017,Tishby2021}.
Yet another important event 
which occurs at much longer time scales
is the step at which the RW completes 
visiting all the nodes in the network.
The time at which this happens is called the cover-time, which scales
like $t \sim N \ln N$
\cite{Cooper2005}.
This means that on average an RW visits each node $\ln N$ times before
it completes visiting all the nodes in the network at least once.

It is interesting to compare the results obtained in this paper for 
RRGs in the infinite network limit with the corresponding results
for regular lattices with the same coordination numbers.
For example, the coordination number of a hypercubic lattice in $d$
dimensions is $2d$. Thus, in terms of the connectivity it is
analogous to an RRG of degree $c=2d$.
An RW on an infinite one dimensional lattice returns to the initial
site with probability $P({\rm R})=1$.
In fact, this result can be obtained by inserting $c=2$ in Eq. (\ref{eq:Pret2}) above.
The distribution of first return times 
of an RW on a one dimensional lattice is given by

\begin{equation}
P(T_{\rm FR}=t) = 
\left\{
\begin{array}{ll}
\frac{ (t-2)! }{ 2^{t-1} \left( \frac{t}{2} - 1 \right)! \left( \frac{t}{2} \right)! }
&   \ \ \ \ \ \   t   \ \    {\rm even} \\
0 & \ \ \ \ \ \  t   \ \ \  {\rm odd}.
\end{array}
\right.  
\label{eq:PTFRt1D}
\end{equation}

\noindent
This result is obtained 
by inserting $c=2$ in
Eq. (\ref{eq:PTFRt2}) above.
The mean first return time of an RW on a one dimensional lattice
diverges. This is consistent with Eq. (\ref{eq:ETFRet}) above, whose right
hand side diverges for $c=2$.

An RW on a two dimensional square lattice also returns to the initial
site with probability $P({\rm R})=1$.
In contrast, for RWs on hypercubic lattices of dimensions $d \ge 3$ the
probability to return to the initial site is $P({\rm R}) < 1$
\cite{Polya1921}.
As shown above, an RW on an RRG of degree $c \ge 3$ and infinite size
returns to the initial node with probability $P({\rm R})=1/(c-1)< 1$.
This means that RWs on RRGs behave qualitatively like RWs on
regular lattices of dimension $d \ge 3$.
For an RW on a hypercubic lattice of dimension $d \ge 3$, the probability of return
to the initial site is given by  
\cite{Hughes1996,Finch2003,Grimmett2018}
 
\begin{equation}
P({\rm R}) = 
1 -
\left\{ \int_{0}^{\infty} e^{-t} \left[ I_0 \left( \frac{t}{d} \right) \right]^d dt \right\}^{-1},
\label{eq:PRlattice}
\end{equation}

\noindent
where $I_0(x)$ is the modified Bessel function
\cite{Olver2010}.
In Appendix A we evaluate the right hand side of Eq. (\ref{eq:PRlattice})
as a Taylor expansion in powers of $1/d$.
We obtain

\begin{equation}
P({\rm R}) =
\frac{1}{2d} + \frac{1}{2d^2} + \mathcal{O} \left( \frac{1}{d^3} \right).
\end{equation}

\noindent
The return probability of an RW on an RRG 
with the same coordination number can be obtained by inserting $c=2d$
in Eq. (\ref{eq:Pret2}). It yields

\begin{equation}
P({\rm R}) = \frac{1}{2d-1}.
\label{eq:PR2d}
\end{equation}

\noindent
Writing Eq. (\ref{eq:PR2d}) as a power series in $1/d$, we obtain

\begin{equation}
P({\rm R}) =
\frac{1}{2d} + \frac{1}{4d^2} + \mathcal{O} \left( \frac{1}{d^3} \right).
\end{equation}

\noindent
This means that to leading order in $1/d$ the return probability of an RW
on a high dimensional hypercubic lattice coincides with the return probability
of an RW on an RRG with the same coordination number.
However, the subleading terms are not the same.
For a $d$-dimensional hypercubic lattice it is $1/(2d^2)$ while for the corresponding
RRG it is $1/(4d^2)$.
The difference is due to the contribution of the shortest cycles in the hypercubic lattice,
whose length is $\ell=4$. 
This is due to the fact that the probability of an RW on a hypercubic lattice to return to
the initial site at time $t=4$ via such cycle is 

\begin{equation}
P(T_{\rm FR}=4)= \frac{1}{4 d^2} + \mathcal{O} \left( \frac{1}{d^3} \right).
\end{equation}

\begin{figure}
\centerline{
\includegraphics[width=7.2cm]{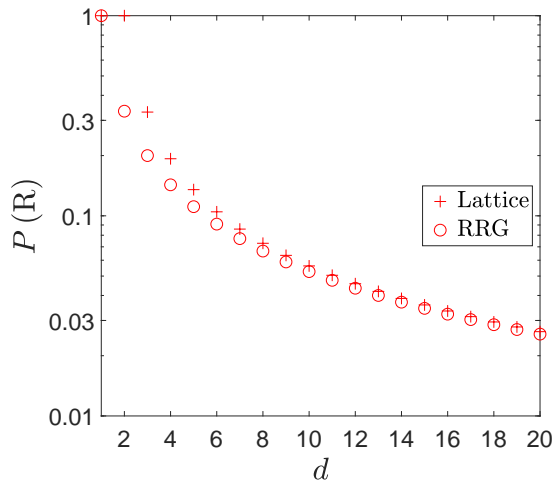}
}
\caption{
The return probability $P({\rm R})$ of an RW on a $d$-dimensional hypercubic lattice   
($+$ symbols) as a function of $d$, obtained from Eq. (\ref{eq:PRlattice}). 
These probabilities are compared to the return probabilities
of RWs on RRGs with the same coordination numbers
$c=2d$ (circles), given by Eq. (\ref{eq:PR2d}).
It is found that the two curves coincide for $d=1$.
For $d \ge 2$ the return probability on a $d$-dimensional hypercubic lattice
is larger than on the corresponding RRG. 
This is due to the existence of short cycles in the hypercubic lattice,
which provide additional channels for the RW to return to the initial node.
As $d$ is increased the contribution of these short cycles diminishes and the
two curves converge towards each other.
}
\label{fig:9}
\end{figure}

In Fig. \ref{fig:9} we present the return probability $P({\rm R})$ 
of an RW on a $d$-dimensional hypercubic lattice   
($+$ symbols) as a function of $d$, obtained from Eq. (\ref{eq:PRlattice}). 
The results are compared to the return probability
of an RW on an RRG with the same coordination number (circles), given by Eq. (\ref{eq:PR2d}).
It is found that as $d$ is increased the two curves converge towards each other.
This means that in the limit of high dimensions the return probabilities of RWs on
hypercubic lattices coincide with those of the corresponding RRGs.
Thus, Eq. (\ref{eq:PR2d}) provides a simple asymptotic expression for Eq. (\ref{eq:PRlattice}) in
the large $d$ limit.
Also note that the return probabilities coincide for $d=1$.
This is typical of the Bethe-Peierls approximation that 
recovers the exact one dimensional result as well as the results in 
high dimensions
\cite{Pathria2011,Plischke2006}.

Another important quantity that is interesting to compare between RRGs and regular lattices
is $\langle S \rangle_t$.
In the case of $d$-dimensional hypercubic lattices, it was shown that
$\langle S \rangle_t \sim \sqrt{t}$ for $d=1$,
$\langle S \rangle_t \sim t/\ln t$ for $d=2$,
while for $d \ge 3$ 
$\langle S \rangle_t = [1-P({\rm R})] t$,
where $P({\rm R})$ is given by Eq. (\ref{eq:PRlattice})
\cite{Montroll1965}.
Thus, the results obtained for $\langle S \rangle_t$ in RRGs
resemble the corresponding results on regular lattices 
of dimensions $d \ge 3$.

After returning to the initial node $i$ for the first time, the RW continues
its trajectory and may return to $i$ again at a later time. 
In the limit of $N \rightarrow \infty$ the RW is transient and it thus
returns to $i$ only a finite number of times.
The distribution of the number of times an RW on an infinite RRG returns to $i$
is given by

\begin{equation}
P(N_{\rm visits}=n) = \frac{c-2}{c-1} \left( \frac{1}{c-1} \right)^n,
\end{equation}

\noindent
where $n=0,1,2,\dots$.
This is a geometric distribution whose moment generating function is given by

\begin{equation}
M(x) = \frac{c-2}{c-1-e^x}.
\end{equation}

\noindent
The cumulant distribution function of $P(N_{\rm visits}=n)$
is given by

\begin{equation}
K(x) = \ln M(x) = \ln \left( \frac{c-2}{c-1-e^x} \right).
\end{equation}

\noindent
Expanding $K(x)$ to second order in $x$, we obtain the mean and variance
of $P(N_{\rm visits}=n)$, which are given by

\begin{equation}
\mathbb{E}[N_{\rm visits}] = \frac{1}{c-2},
\end{equation}

\noindent
and

\begin{equation}
{\rm Var}(N_{\rm visits}) = \frac{c-1}{(c-2)^2},
\end{equation}

\noindent
respectively.
Thus, the expected number of visits decreases as $c$ is increased.

\section{Summary}

We presented analytical results for 
the distribution of first return times 
of RWs on RRGs.  
In the analysis we distinguished between
the scenario in which the RW returns
to the initial node $i$ by retroceding its own steps and
the scenario in which it does not retrocede its steps 
on the way back to $i$.
We calculated the
conditional distributions of first return times,
$P(T_{\rm FR}=t | {\rm RETRO})$
and
$P(T_{\rm FR}=t | \lnot {\rm RETRO})$,
in the retroceding and non-retroceding scenarios, respectively.
We also calculated the mean and the variance of each one of the two conditional distributions.
It was found that the distribution of first return times in the retroceding 
scenario is narrow and its mean is of order $1$.
In contrast, the distribution of first return times in the non-retroceding
scenario is broad and its mean is $\simeq N$.
Thus, there is a clear separation of time scales between the two scenarios.
We combined the results of the two scenarios and obtained
the overall distribution of first return times
$P(T_{\rm FR} = t)$
of RWs on RRGs of a finite size.
The retroceding scenario is retained in the infinite network limit,
while the non-retroceding scenario exists only in finite networks.
It was also found that in the infinite network limit the trajectories of RWs on RRGs
are transient in the sense that they return to the initial node with probability $<1$.
In this sense they resemble the trajectories of RWs on regular lattices of dimensions
$d \ge 3$.

This work was supported by the Israel Science Foundation grant no. 
1682/18.

\appendix

\section{Asymptotic expansion of P\'olya's constant}

The probability that an RW on a $d$-dimensional hypercubic lattice will
return to the initial site is given by

\begin{equation}
P({\rm R}) = 1 - \frac{1}{U(d)},
\label{eq:P(R)}
\end{equation}

\noindent
where

\begin{equation}
U(d) = \int_{0}^{\infty} e^{-t} \left[ I_0 \left( \frac{t}{d} \right) \right]^d dt,
\end{equation}

\noindent
is P\'olya's constant and $I_0(x)$ is the modified Bessel function of the first kind
\cite{Olver2010}.
Changing variables from $t$ to $s=t/d$, one can express $U(d)$ in the form

\begin{equation}
U(d) = d \int_{0}^{\infty} e^{-ds} \left[ I_0 \left( s \right) \right]^d ds,
\label{eq:Udint}
\end{equation}

\noindent
which is $d$ times the Laplace transform of $[I_0(s)]^d$.
In order to carry out the integration, we use the Taylor expansion of
$[I_0(s)]^d$ around $s=0$ and integrate it term by term.
The expansion is given by
\cite{Bender2003,Baricz2010,Moll2014}

\begin{equation}
[I_0(s)]^d = 1 + d \sum_{k=0}^{\infty}
\frac{ P_k(d) }{ \left[ (k+1)! \right]^2 }
\left( \frac{s}{2} \right)^{2(k+1)},
\label{eq:I0}
\end{equation}

\noindent
where $P_k(d)$ are the Bender-Brody-Meister polynomials,
which are expressed in powers of $d$ and have integer coefficients.
The lowest order polynomials and the recursion equation for generating
higher order polynomials appear in Ref. \cite{Bender2003}.
Inserting $[I_0(s)]^d$ from Eq. (\ref{eq:I0}) into Eq. (\ref{eq:Udint})
and carrying out the integration term by term, we obtain

\begin{equation}
U(d) = 1 + d \sum_{k=1}^{\infty}
\binom{2k}{k} \frac{P_{k-1}(d)}{(2d)^{2k}}.
\end{equation}

\noindent
Using the lowest order polynomials $P_0(d),\dots,P_6(d)$,
which are given explicitly in Ref. \cite{Bender2003}, we obtain

\begin{equation}
U(d) = 1 + \frac{1}{2d} + \frac{3}{4d^2} + \frac{3}{2d^3}
+ \frac{15}{4d^4} + \frac{355}{32d^5} + \frac{595}{16d^6} + \frac{8715}{64d^7} 
+ \mathcal{O} \left( \frac{1}{d^8} \right).
\label{eq:U(d)}
\end{equation}

\noindent
Inserting $U(d)$ from Eq. (\ref{eq:U(d)}) into Eq. (\ref{eq:P(R)}), we obtain

\begin{equation}
P({\rm R}) = \frac{1}{2d} + \frac{1}{2d^2} + \frac{7}{8d^3}
+ \frac{35}{16d^4} + \frac{215}{32d^5} + \frac{1501}{64d^6}
+ \frac{5677}{64d^7} + \mathcal{O} \left( \frac{1}{d^8} \right).
\end{equation}

\noindent
This $7$th order expansion in powers of $1/d$ provides an approximation for
Eq. (\ref{eq:P(R)}) whose error is less than $1\%$ for $d \ge 4$ and less than
$0.01 \%$ for $d \ge 10$.

\noappendix

\end{document}